\documentclass[twocolumn,tighten,times]{aastex631}

\newcommand{\mcl}[1]{\textcolor{black}{#1}} 

\begin{document}

\title{Maximum Energy of Particles in Plasmas}

\correspondingauthor{Mitsuo Oka}
\email{moka@berkeley.edu}

\author[0000-0003-2191-1025]{Mitsuo Oka}
\affiliation{Space Sciences Laboratory, University of California Berkeley, 7 Gauss Way, Berkeley, CA 94720, USA}

\author[0000-0002-1040-8769]{Kazuo Makishima}
\affiliation{Department of Physics, The University of Tokyo, 7-3-1 Hongo, Bunkyo-ku,Tokyo 113-0033, Japan}
\affiliation{Kavli Institute for the Physics and Mathematics of the Universe (WPI), The University of Tokyo, 5-1-5 Kashiwa-no-ha, Kashiwa,
Chiba 277-8683, Japan}

\author[0000-0003-2507-9803]{Toshio Terasawa}
\affiliation{Institute for Cosmic Ray Research, The University of Tokyo, 5-1-5 Kashiwa-no-Ha, Kashiwa, Chiba 277-8582, Japan}








\begin{abstract}

Particles are accelerated to very high, non-thermal energies in space, solar, and astrophysical plasma environments. In cosmic ray physics, the {\it Hillas limit} is often used \mcl{as a rough estimate (or the necessary condition) of} the maximum  energy of particles. This limit is based on the concepts of one-shot direct acceleration by a system-wide motional electric field, as well as stochastic and diffusive acceleration in strongly turbulent environments. However, it remains unclear how well this limit explains the actual observed maximum energies of particles. Here we show, based on a systematic review, that the observed maximum energy of particles  --- those in space, solar, astrophysical, and laboratory environments --- often reach the energy predicted by the Hillas limit. We also found several exceptions, such as electrons in solar flares and jet-terminal lobes of radio galaxies, as well as protons in planetary radiation belts, where deviations from this limit occur. We discuss possible causes of such deviations, and we argue in particular that there is a good chance of detecting ultra-high-energy \mcl{($\sim$100 GeV)} solar flare electrons that have not yet been detected. We anticipate that this study will facilitate further interdisciplinary discussions on the maximum energy of particles and the underlying mechanisms of particle acceleration in diverse plasma environments.

\end{abstract}




\section{Introduction} \label{sec:intro}

Particles -- both ions and electrons -- are accelerated to very high non-thermal energies in space, solar, and astrophysical plasma environments \citep[e.g.][and references therein]{BlandfordR_1987, TerasawaT_2001, KruckerS_2008_review, OkaM_2018, OkaM_2023, GuoFan_2024}. 
\mcl{While the mechanism of particle acceleration has been a topic of major debate, a significant question remains regarding the maximum particle energy that can be reached in each plasma environment.}

\mcl{In an effort to constrain potential origins of ultra-high-energy ($>10^{18}$ eV) cosmic rays (UHECRs),  \cite{HillasAM_1984} argued that the size $L_{\rm 0}$ of the essential part of the acceleration region must be greater than $2r_{\rm g}$, where $r_{\rm g}$ is the gyro-radius of UHECRs. If particles are accelerated stochastically, the diffusive motion needs a very large space, much larger than the gyro-radius, so that $L_{\rm 0}>2r_{\rm g}/\beta$, where $\beta=V/c$ and $V$ is  the characteristic velocity of the scattering centers. Based on these considerations, \cite{HillasAM_1984} presented what is now known as the Hillas diagram, which illustrates possible candidates of UHECR sources in the $B - L_{\rm 0}$ parameter space, where $B$ is the magnetic field strength. It was proposed as a way to identify which sources fail to satisfy the requirements for UHECR sources.}

\mcl{Evidently, the arguments can be adapted in more generic discussions of maximum energy of particles in various plasma environments, well beyond the topic of search for UHECR production sites. The gyro-radius of relativistic particles is expressed as  $r_g \sim \varepsilon/cqB$ (SI unit) where $\varepsilon$ is the particle energy, $c$ is the light speed, and $q$ is the particle charge. Then, the condition of $L_{\rm 0} > 2r_{\rm g}$ gives the {\it hard} limit for the maximum particle energy achievable in a plasma environment:
\begin{equation}
\label{eq:hillas-c}
    \varepsilon_{\text{\scriptsize{limit}}} = qcBL
\end{equation}
where we omitted the factor of 2 by introducing the half-width of the acceleration region $L=L_{\rm 0}/2$. A caveat here is that the statement $L > r_{\rm g}$ is a necessary condition, rather than a sufficient condition, for particles to be trapped in the acceleration region.  This is because particles might escape from the system even when $L > r_{\rm g}$.}

\mcl{Hillas's second condition of $L_{\rm 0}>2r_{\rm g}/\beta$ can be rewritten to provide a more stringent limit of the maximum energy of particles
\begin{equation}
\label{eq:hillas-V}
    \varepsilon_{\rm H} = qVBL
\end{equation}
which we hereafter refer to as the {\it Hillas limit}. While \cite{HillasAM_1984} envisioned the stochastic acceleration scenario, the same expression can} be obtained by assuming a one-shot, direct acceleration by the system-wide motional electric field $VB$. In fact, similar forms of Eq.(\ref{eq:hillas-V}) have been widely used in the context of magnetic reconnection where particles can be accelerated by the coherent reconnection electric field \citep[e.g.][]{MatthaeusWH_1984, GoldsteinML_1986, MartensPCH_1988, KliemB_1994}. \cite{MakishimaK_1999} also used such an idea of direct acceleration to arrive independently at the same scaling as Eq.(\ref{eq:hillas-V}), and argued that the predicted $\varepsilon_{\rm H}$ is often close to the highest particle energies $\varepsilon_{\rm obs}$ found  in various plasma environments where \mcl{not only magnetic reconnection but also shocks} could play a major role.

Interestingly, \cite{TerasawaT_2001} obtained \mcl{essentially the same limit as Eq.(\ref{eq:hillas-V})} by assuming more explicitly that particles are accelerated stochastically in a diffusive environment. Using the spatial diffusion coefficient $D$, the size of the acceleration region can be roughly expressed as $\lambda = D/V$ \citep[e.g.][]{BlandfordR_1987}. If we normalize $D$ by the Bohm diffusion coefficient $D_B = (1/3)r_g v$ (where $v$ denotes the particle speed) and introduce $\eta \equiv D/D_B$, we obtain $\lambda = \eta D_B/V = \eta \varepsilon/3qVB$ where we used the relativistic gyro-radius. Then, from the condition  $\lambda \leq L$, we obtain the maximum attainable energy as
\begin{equation}
\label{eq:terasawa}
    \varepsilon_{\text{\scriptsize{diffusive}}} = \frac{3}{\eta} qVBL
\end{equation}
In the strong scattering limit ($\eta=1$), this maximum energy is a factor of 3 higher than $\varepsilon_{\rm H}$, although $\eta$ is likely much larger than unity in solar and space plasmas. In the non-relativistic regime, we can readily obtain $\varepsilon_{\text{\scriptsize{diffusive}}}  = (3/2\eta)qVBL$ which is still comparable to $\varepsilon_{\rm H}$ in the strong scattering limit. Therefore, \mcl{we should keep in mind that Eq.(\ref{eq:hillas-V}) implies not only  one-shot direct acceleration scenario but also the stochastic acceleration scenario.}


\mcl{A crucial point to note is that Eq.(\ref{eq:hillas-V}) remains a necessary, but not a sufficient, condition \citep[e.g.][]{PtitsynaKV_2010, KoteraK_2011, AlvesBatistaR_2019, MatthewsJH_2020}. More stringent conditions may arise from additional processes. In fact, \cite{HillasAM_1984} already discussed, for example, possible compositions of UHECRs, associated transport processes, and energy losses due to synchrotron radiation. More up-to-date descriptions of the limitation factors for the maximum particle energy are well summarized by \cite{MatthewsJH_2020}, particularly their Figure 5. In other words, with the understanding that the Hillas limit is a ballpark estimate, the cosmic ray physics community has sought more restrictive conditions in each plasma environment. } 

\mcl{In contrast, \cite{MakishimaK_1999} and \cite{TerasawaT_2001, TerasawaT_2011} took a different approach, aiming to validate the Hillas limit itself by comparing it with  observations of high-energy particles and photons. They suggested that particle energies do reach the Hillas limit in various acceleration regions, even in solar and space plasma environments. \mcl{On the other hand}, a recent study suggests that the Hillas limit may overestimate observed maximum energies in many plasma environments \citep{ChienA_2023}. Regardless, these studies did not fully account for additional loss processes considered in the context of UHECR studies.}

\mcl{This Paper presents a novel study that integrates both approaches. Leveraging more recent observations of space, solar, astrophysical, and laboratory plasmas, we first attempt to validate the Hillas limit itself. However, when interpreting the results,  we consider radiative energy losses as well as the scattering intensity, parameterized by $\eta$. A large value of $\eta$ represents weak scattering, less effective confinement of particles in the acceleration region, and therefore a condition more restrictive than the Hillas limit.} We also treat ions and electrons separately. This is because, although the Hillas limit does not depend on the particle mass, ions and electrons could experience different acceleration and energy loss processes. Additionally, we try to use consistent definitions for $V$, $B$, and $L$ for all plasma environments. \mcl{With these new approaches, we have achieved a much better understanding of the maximum energy of particles in plasmas.}

We first describe the methodology including the definitions of the $V,B,L$ parameters (Section \ref{sec:method}), followed by detailed descriptions of how we collected and compiled the parameter values, as well as the observed maximum energies of particles (Section \ref{sec:data}). Then, we compare the observations and predictions in graphs (Section \ref{sec:results}), discuss the results (Section \ref{sec:discussion}), and conclude at the end (Section \ref{sec:conclusion}).

\section{Methodology}\label{sec:method}

\begin{figure*}
\plotone{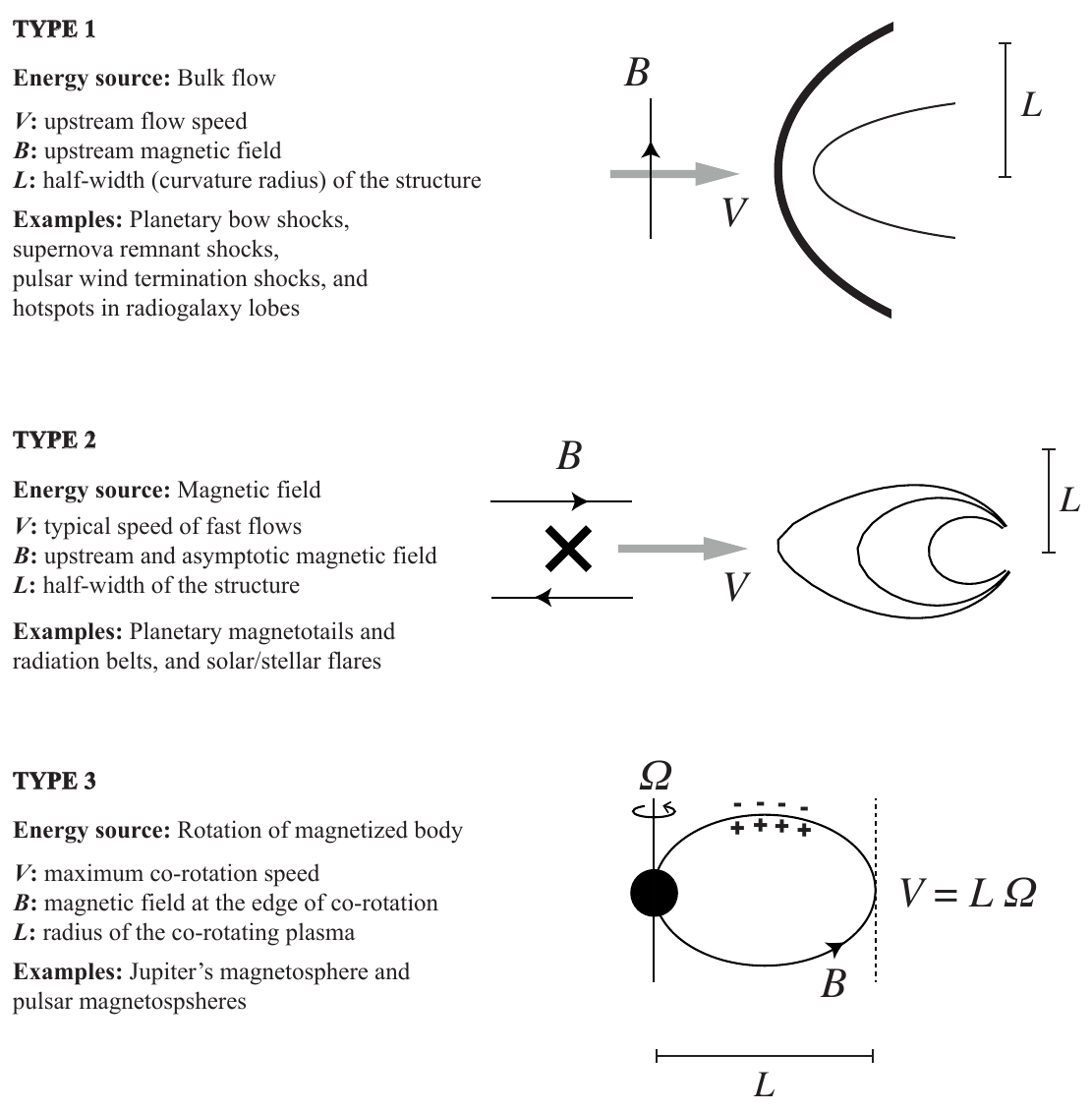}
\caption{\mcl{Schematic illustration of different types of plasma environments. Annotated on the left of each illustration are (1) the primary source of energy, (2) our definitions of $V$, $B$, and $L$, and (3) example plasma environments. This categorization is not meant to restrict the acceleration mechanism. For example, in the Type 2 environment,  particles may be accelerated not only by magnetic reconnection but also by the termination shock formed by the dissipation of reconnection jets at the apex of a flaring loop. }  For more details, see Section \ref{sec:method} and Table \ref{tab}.
\label{fig:illustration}}
\end{figure*}


We conducted a systematic review of the literature to deduce the key parameters ($V$, $B$, and $L$), predict the maximum particle energies $\varepsilon_{\rm H}$ via Eq.(\ref{eq:hillas-V}), and compare these predictions with the highest, observed energies of protons ($\varepsilon_{\rm obs, p}$) and electrons ($\varepsilon_{\rm obs, e}$).  We selected plasma environments from which we can confidently derive these key parameters as well as  $\varepsilon_{\rm obs}$. While heavy ions are also observed in some cases, we focus on protons in this study for simplicity.

There are three caveats when conducting this study. The first is that, in many cases, the observed maximum energies ($\varepsilon_{\rm obs}$) represent only a lower limit of the possible maximum particle energy.  This is because the higher-energy cutoff in power-law spectra is often unclear, particularly in solar and space plasmas. Power-laws with exponential cut-offs \citep{EllisonDC_1985, BandD_1993, LiuZixuan_2020} can be found in, for example, Solar Energetic Particle (SEP) events \citep[e.g.][]{EllisonDC_1985, MewaldtRA_2012a} and crossings of Earth's bow shock \citep[e.g.][]{OkaM_2006, AmanoT_2020}.  However, observed energy spectra do not always exhibit such a cutoff feature in many plasma environments. The $\varepsilon_{\rm obs}$ values are typically derived from the highest energy channel of instruments that registered significant particle or photon counts. These values may be updated in the future with improved measurements that feature a larger energy range and higher sensitivity (or a larger dynamic range).  Additionally, in the case of astrophysical objects, radiative energy losses could reduce the actual maximum energy of particles \citep{PtitsynaKV_2010}. Therefore, $\varepsilon_{\rm obs}$ values that are lower than  the Hillas limit do not automatically invalidate Eq.(\ref{eq:hillas-V}).  Similarly, $\varepsilon_{\rm obs}$ values that are comparable to the Hillas limit $\varepsilon_{\rm H}$ do not necessarily confirm its validity, as the actual maximum energy could be exceeding the Hillas limit. However, $\varepsilon_{\rm obs} \sim \varepsilon_{\rm H}$ provides a strong evidence for particle acceleration mechanisms capable of reaching the Hillas limit energies.

Second, while direct {\it in-situ} measurements are available for space plasmas, remote-sensing measurements provides the key for diagnosing astrophysical plasmas. The observed maximum energy of photons can be identified, but it needs to be converted to the corresponding particle energy after knowing the emission mechanism and relevant parameters. For example, synchrotron radiation depends on the magnetic field $B$, and so we need to treat the estimation of both $B$ and $\varepsilon_{\rm obs}$ as a set of simultaneous equations.

Finally, $V$, $B$, and $L$ should be defined in a way that is as  consistent as possible across  different plasma environments. As described in previous section, the Hillas limit can imply both one-shot direct acceleration and stochastic acceleration. Also, it does not necessarily distinguish shocks and magnetic reconnection. Therefore, we prefer to have a definition that does not explicitly depend on a specific acceleration mechanism. Nevertheless, the plasma condition and geometry differ greatly from acceleration site to site. To define $V$, $B$, and $L$, it is therefore still convenient to categorize plasma environments based on the type of primary source of energy. Figure \ref{fig:illustration} illustrates the categorization and our definitions of $V$, $B$, and $L$. 

The first type of plasma environment (Fig.\ref{fig:illustration}, top) has the upstream bulk flow as the primary source of energy. Examples include planetary bow shocks and the termination shocks of the heliospheric solar wind,  pulsar winds, and jets in radio galaxies. This configuration does not necessarily mean particles are accelerated by shocks only. It is certainly possible that particles are accelerated by magnetic reconnection that occurs, for example, in the current sheets embedded in solar and pulsar winds \citep[e.g.][]{DrakeJF_2010, ZankGP_2014, ZankGP_2015, LuYingchao_2021}.  We define $V$ as the bulk flow speed on the upstream side and $B$ as the  strength of the magnetic field carried by the flow.  In astrophysical non-thermal sources, a magnetic field strength may be  obtained through observations of emissions from the post-shock region, but we regard such a value as the magnetic field on the downstream side. Upon application of the Hillas limit, we reduce such a post-shock value by dividing by the shock compression ratio to derive the upstream value. The spatial scale $L$ is approximated by the curvature radius of the shock front. In case of shell-type supernova remnants, we use the radius of the shell instead of the thickness of the shell, although we also provide thickness-based estimates for reference. In case of a radio galaxy,  twin jets from its nucleus terminate via shocks, to form gigantic radio lobes filled with energetic particles. Although the shock front itself may not be spatially resolved via imaging \mcl{in any wavelengths}, we can still use the radius of the so-called hot-spots that sit at lobe centers which is clearly visible \mcl{in radio images}. In terrestrial/planetary bow shocks, the curvature radius could roughly correspond to the half-width of the obstruction, i.e., the magnetotail. It should also be roughly comparable to the standoff distance between the nose of the shock and the central planet.

The second type of plasma setting (Fig.\ref{fig:illustration}, center) has  accumulated magnetic flux in the upstream region as the primary source of energy. 
In such a situation, the inflowing anti-parallel fields reconnect to produce bi-directional jets, one of which impinges on quasi-stationary magnetic fields anchored to some thick materials. Examples include solar flares and Earth's magnetosphere. This categorization does not necessarily mean we exclude the possibility of other particle acceleration mechanisms. In solar flares, it is certainly possible that particles are accelerated by the termination shocks located at or above the top of the magnetic loop on the downstream side \citep[e.g.][]{MasudaS_1994, TsunetaNaito_1998, ChenB_2015}. We define $B$ as the characteristic (asymptotic) magnetic field on the upstream side of reconnection, such as the lobe magnetic field in Earth's magnetotail and ambient magnetic field in the solar corona. We define $V$ as the typical speed of the fast flows in the system to obtain an estimate of the maximum attainable energy in the system. It should be comparable to the reconnection outflow speed or the Alfv\'{e}n speed $V_A$.   Planetary radiation-belts also falls into this category. Although the solar wind dynamic pressure could be contributing to a global-scale compression of the inner magnetosphere and hence particle energization in the radiation belts, observations also indicate that a substantial amount of particle and energy flux is `injected' into the inner magnetosphere from the magnetotail. 

The third type of plasma configuration involves a rapidly rotating, strongly magnetized planet or star, with the electric potential induced by the co-rotating magnetic field. Examples include the magnetospheres of  Jupiter and pulsars such as the Crab Pulsar. While the precise mechanisms of particle acceleration remain unclear in such environments, a thin layer of charge separation which we refer to as a double-layer is often considered. As in the former two categories, the acceleration could also be driven by other mechanisms, such as magnetic reconnection and pitch angle scattering by waves. We define $V$ as the maximum co-rotation speed, $L$ as the radial distance from the central object to the location of the maximum co-rotation speed, and $B$ is the typical magnetic field near the radial distance of $L$.

\begin{deluxetable*}{clrrrrrr}\label{tab}
    \tablecaption{The Hillas limit and the observed maximum energy of particles in various plasma environments.}
    \tablehead{\multicolumn2l{Plasma environment} & \colhead{Flow speed} & \colhead{Magnetic field} & \colhead{System Size} & \colhead{Hillas limit}  & \multicolumn2c{Observations}\\
    \multicolumn2l{Type$\quad$Name} & \colhead{$V$ (km s$^{-1}$)} & \colhead{$B$ (nT)} & \colhead{$L$ (m)} & \colhead{$\varepsilon_{\rm H}$} & \colhead{$\varepsilon_{\rm obs, p}$} & \colhead{$\varepsilon_{\rm obs, e}$}
    }
    \startdata
2&Earth's magnetotail&$300 - 1000$&$15 - 25$&$(9.6 - 16) \times 10^{7}$&0.4 - 4 MeV&1 MeV&2 MeV\\
2&Earth's radiation belt&$300 - 1000$&$15 - 25$&$(9.6 - 16) \times 10^{7}$&0.4 - 4 MeV&800 MeV&15 MeV\\
1&Earth's bow shock&$400 - 600$&$5 - 8$&$(9.6 - 16) \times 10^{7}$&190 - 760 keV&300 keV&300 keV\\
2&Mercury's magnetosphere&$300 - 400$&$40 - 50$&$(4.9 - 9.8) \times 10^{6}$&59 - 200 keV&550 keV&300 keV\\
3&Jupiter's radiation belt&$400 - 800$&$5 - 15$&$(5.2 - 8.2) \times 10^{9}$&10 - 98 MeV&80 MeV&100 MeV\\
3&Saturn's radiation belt&$200 - 500$&$5 - 10$&$(1.2 - 2.9) \times 10^{9}$&1 - 15 MeV&300 MeV&2 MeV\\
1&Saturn's bow shock&$300 - 600$&$0.2 - 1.0$&$(2.9 - 3.5) \times 10^{9}$&0.2 - 2 MeV&$-$&700 keV\\
1&Heliosphere (ACR)&$250 - 350$&$0.02 - 0.1$&$(1.2 - 1.8) \times 10^{13}$&60 - 630 MeV&100 MeV&$-$\\\hline
2&Solar flares&$1000 - 3000$&$(5 - 50) \times 10^{6}$&$(1.0 - 3.0) \times 10^{7}$&0.05 - 5 TeV&30 GeV&45 MeV\\
1&CME (shock)&$1000 - 4000$&$300 - 10000$&$(3.5 - 10) \times 10^{9}$&1 - 420 GeV&30 GeV&$-$\\\hline
1&SN 1006&$2700 - 4600$&$0.1 - 0.5$&$2.7 \times 10^{17}$&73 - 620 TeV&100 TeV&10 TeV\\
1&RX J1713.7-3946&$3900$&$0.1 - 1.0$&$1.4 \times 10^{17}$&53 - 530 TeV&100 TeV&100 TeV\\
1&Crab Nebula&$3 \times 10^{{{5}}}$&$1 - 4$&$4.3 \times 10^{15}$&2 - 6 PeV&$-$&2 PeV\\
3&Crab Pulsar&$3 \times 10^{{{5}}}$&$1.0 \times 10^{11}$&$1.5 \times 10^{6}$&45 PeV&$-$&15 TeV\\
1&Cygnus A&$72000 - 90000$&$9 - 10$&$(3.4 - 4.6) \times 10^{19}$&22 - 42 EeV&$-$&30 GeV\\\hline
1&Laser plasma (shock p)&$1500$&$2.0 \times 10^{10}$&$(3.0 - 4.0) \times 10^{-3}$&90 - 120 keV&80 keV&$-$\\
1&Laser plasma (shock e-)&$1000$&$1.0 \times 10^{11}$&$5.0 \times 10^{-3}$&500 keV&$-$&500 keV\\
2&Laser plasma (mrx e-)&$500 - 1100$&$5.0 \times 10^{10}$&$1.0 \times 10^{-3}$&25 - 55 keV&$-$&$40-70$ keV\\
    \enddata
    \tablecomments{ The first column shows the type of each environment  (either 1, 2, or 3). See Section \ref{sec:method} for more details. The plasma environments are listed in the order of appearance in Section \ref{sec:data}. The average radii of the planets are as follows: Earth $R_{\rm E} \sim$  6.371 $\times 10^6$ m, Mercury $R_{\rm M} \sim$ 2.440 $\times 10^6$ m,  Jupiter $R_{\rm J} \sim$ 6.9911 $\times 10^7$ m, and Saturn $R_{\rm S} \sim$ 5.8232 $\times 10^7$ m. The average solar radius is $\sim$6.957 $\times 10^8$ m.  For unit conversions,  1 nT = 10 $\mu$G.}
\end{deluxetable*}

\section{Data Collection\label{sec:data}}

Based on the methodology described above, we searched the literature and collected  data, as summarized in Table \ref{tab}. In this section, we describe the details of the collected data.

\subsection{Space Plasmas}\label{sec:space}

Let us start from Earth's magnetotail because it has been extensively investigated over the past decades. To estimate $\varepsilon_{\rm H}$, we consider `near-Earth' reconnection that takes place in the down-tail distances of $\sim 15 - 25 R_{\rm E}$ where $R_{\rm E}$ is Earth's radius  \citep[e.g.][]{MiyashitaY_2009}. In this region, the lobe magnetic field is $B = 15 - 25$ nT \citep{FairfieldDH_1996} and the half-width of the magnetotail is $L = 15-25 R_{\rm E}$ or $(9.6 - 16) \times 10^7$ m \citep{FairfieldDH_1971}. For the speed $V$, we use the typical speed of bursty bulk flows $V = 300 - 1000$ km s$^{-1}$ in the central plasma sheet \citep[e.g.][]{AngelopoulosV_1992, ZhangLQ_2016}. While the flow channel appears to be much narrower than the full width of the magnetotail, on the order of a few $R_{\rm E}$ \citep[e.g.][]{AngelopoulosV_1996}, we adhere to our definition described in Section \ref{sec:method} and use the magnetotail half-width for $L$. Consequently, we obtain $\varepsilon_{\rm H} = 0.4 - 4.0$ MeV, while earlier observations have reported up to $\sim 1$ MeV for both ions and electrons \citep[e.g.][]{TerasawaT_1976, KrimigisSM_1979, ChristonSP_1988, ChristonSP_1989, ChristonSP_1991}. A recent study also reported a detection of $\sim$ 2 MeV electrons that were precipitating into Earth's atmosphere from the magnetotail \citep{ArtemyevAV_2024}. It is worth noting that earlier studies already argued that the direct acceleration by the reconnection electric field does not explain $\sim$ MeV particles \citep[e.g.][]{TerasawaT_1976, KrimigisSM_1979}.  In fact, the polar cap potential, which is formed when the dayside magnetopause reconnection enables the solar wind to drive ionospheric convection, is known to saturate only at $\sim$200 kV \citep{HairstonMR_2005, ShepherdSG_2007}.

Earth's radiation belt is also a distinct region of particle acceleration, but the values of $B$ vary largely and the values of $V$ can be fluctuating around $\sim$0 due to its oscillatory motion, making it difficult to choose a characteristic value of $B$ and $V$. Therefore, we use the characteristic values in the magnetotail, as also described in Section \ref{sec:method}. This is because, while electrons can be accelerated locally inside the radiation belt, a substantial fraction of their energy originates from the magnetotail in the forms of waves and moderately accelerated seed particles \citep[e.g.][and references therein]{JaynesAN_2015, SorathiaK_2018, TurnerDL_2021_source}.  The highest energies of observed particles are $\sim$800 MeV for protons \citep{MazurJE_2023} and $\sim$15 MeV (or possibly higher) for electrons \citep{BlakeJB_1992, VampolaAL_1992}. However, such high-energy protons are likely to be supplied by galactic cosmic rays (GCRs). It has been known that GCR particles knock neutrons out from the atmospheric atoms and such neutrons decay to produce protons \citep[e.g.][]{SingerSF_1958a, SingerSF_1958b, HessWN_1959, LiYuXuan_2023}. This is called Cosmic Ray Albedo Neutron Decay (CRAND). More details of CRAND are discussed in Section \ref{sec:CRAND}.

For Earth's bow shock, we consider the typical values of the `fast' solar wind  at 1 AU  \citep[e.g.][]{LarroderaC_2020}, i.e., $B$ = 5 - 8 nT and $V$ = 400 - 600 km s$^{-1}$. The system size $L$ should be roughly the same as that of the magnetotail, $L = 15 - 25 R_{\rm E}$, or $(9.6-16)\times 10^7$ m. As a result, we expect $\varepsilon_{\rm H} \sim 190 - 760$ keV, while observations in the shock upstream region find energetic protons of up to $\sim$300 keV \citep[e.g.][]{ScholerM_1981, TurnerDL_2018, TrattnerKJ_2023} and electrons of similar energies \citep[e.g.][]{WilsonLB_2016}, although these energetic protons my be partially contributed by those from the magnetosphere \citep{ScholerM_1981}. A recent observation of hot flow anomalies --- transient concentrations of shock-reflected, supra-thermal ions that occur when a certain structures in the solar wind collide with the bow shock ---  reported energizations of ions \citep{TurnerDL_2018}. In their report, the energy spectrum of H$^{+}$, He$^{2+}$, and O$^{6+}$ extended up to $\sim100$, $\sim300$, and $\sim600$ keV, while showing a peak at $\sim$30, $\sim$120, and $\sim$300 keV, respectively. While the authors successfully interpreted the peak energies by a Fermi-type acceleration `trap' between two converging magnetic mirrors, such a dependence on charge state could also be attributed to Eq.(\ref{eq:hillas-V}).

In Mercury's magnetosphere with its half-width being $L \sim 2 - 4 R_{\rm M}$ where $R_{\rm M}$ is Mercury's radius or $(4.9 - 9.8) \times 10^6$ m \citep{WinslowRM_2013}, protons of up to $\sim$ 550 keV \citep{SimpsonJA_1974} and electrons of up to $\sim$ 300 keV \citep{SimpsonJA_1974, HoG_2011_Science, HoG_2012, LawrenceDJ_2015} have been detected. The magneotail reconnection appears to occur around $\sim3 R_{\rm M}$ down-tail, and the lobe magnetic field in such a region is $\sim40 - 50$ nT \citep{PohG_2017, BowersCF_2024}. The bulk flow speed $V$ has been neither directly nor routinely measured, but we assume $300 - 400$ km s$^{-1}$ based on flow speeds associated with dipolarization \citep{DeweyRM_2018} and flux ropes (FTEs) \citep{SlavinJA_2012}. This is comparable to the estimated thermal speed of ions \citep{RainesJM_2011}. As a result, we obtain $\varepsilon_{\rm H} = 59 - 200$ keV.

Jupiter's plasma environment is very different from those of rocky planets such as Earth and Mercury. Its dynamics is largely driven by the rapid planetary rotation, assisted by the significant internal mass loading from the volcanic moon Io \citep[e.g.][and references therein]{JackmanCM_2014, GuoRuilong_2024}. A key picture here is that, at a certain distance from the central planet, the magnetic field is stretched outward and the co-rotation breaks down due to magnetic reconnection \citep{VasyliunasVM_1983}. This is consistent with Type 3 illustrated in Figure \ref{fig:illustration}, and we use the typical radial distance at which the magnetic reconnection takes place, i.e., $L \sim 90-140 R_{\rm J}$ where $R_{\rm J}$ is Jupiter's radius, or $(5.2 - 8.2)\times 10^9$ m \citep{VogtMF_2010, VogtMF_2020}. The lobe magnetic field at such radial distances is $5 - 15$ nT \citep{KivelsonM_2002} and the typical speed of fast plasma flows is  $V \sim 400 - 800$ km s$^{-1}$ \citep{KronbergEA_2008}. As a result, we obtain $\varepsilon_{\rm H} = 10 - 98$ MeV. For observation, synchrotron emissions from $\gtrsim$50 MeV electrons in the radiation belt was first reported by \cite{BoltonSJ_2002}, and it was suggested that there is a maximum cutoff at $\sim$100 MeV \citep{dePaterI_2003}. We also add that non-thermal bremsstrahlung X-rays of up to 20 keV have been detected from the polar (auroral) regions of Jupiter \citep{MoriK_2022}. For protons, the maximum observed energy of $\sim$80 MeV was recorded near the main ring of Jupiter \citep{FischerHM_1996, KollmannP_2017}. Notably, heavy ions of $\gtrsim$125 MeV are also detected based on ionization signatures in the star camera onboard the Juno spacecraft \citep{BeckerHN_2021}. 


Saturn's plasma environment is in many ways intermediate between those of Earth and Jupiter \citep{JackmanCM_2014}. However, the effect of planetary rotation appears to be dominant \citep[e.g.][]{GuoRuilong_2024} and so we consider Saturn's magnetosphere as one of the third type described in Section \ref{sec:method}. On average, the X-line is located at distances of $20 - 30 R_{\rm S}$ where $R_{\rm S}$ is Saturn's radius, but it is highly variable and could be located as far out as $\sim50 R_{\rm S}$ \citep{McAndrewsHJ_2009,SmithAW_2016}. Thus, we adopt the spatial scale of $L = 20 - 50 R_{\rm S}$, or $(1.2 - 2.9)\times 10^9$ m. The lobe magnetic field in such locations is $5 - 10$ nT \citep{JackmanCM_2014}. With the possible flow speeds $V = 200 - 500$ km s$^{-1}$ \citep[e.g.][]{HillTW_2008, McAndrewsHJ_2009}, we obtain $\varepsilon_{\rm H} = 1 - 15$ MeV. In reality, protons with energies up to at least $\sim300$ MeV \citep[e.g.][]{RoussosE_2018, KruppN_2018} and electrons up to at least $\sim$2 MeV \citep[e.g.][]{ParanicasC_2010, RoussosE_2014} have been observed in Saturn's radiation belt. Here, the high-energy protons are likely CRAND protons that also exist in Earth's radiation belt. 

At Saturn's bow shock, electrons up to $\sim$700 keV were detected when the solar wind Alfv\'{e}n Mach number $M_A$ was unusually high $\sim$100 \citep{MastersA_2013}. Based on the solar wind parameters at Saturn, i.e., $B = 0.2 - 1.0$ nT and $V = 300 - 600$ km s$^{-1}$ \citep[e.g.][]{MastersA_2013, ThomsenMF_2019, GershmanD_2024}, as well as the half-width of the magnetosphere $L = 50 - 60 R_{\rm S}$, or $(2.9 - 3.5)\times 10^9$ m \citep[e.g.][]{McAndrewsHJ_2009, JackmanCM_2019}, we estimate $\varepsilon_{p} = 0.2 - 2$ MeV. 

For the heliosphere, we consider Anomalous Cosmic Rays (ACRs) that appear as an excess flux in the lowest energy range of the galactic cosmic ray spectra \citep[e.g.][and references therein]{GiacaloneJ_2022}.
In today's standard scenario, neutral particles of interstellar origin are ionized deep inside the heliosphere and then `picked-up' by the solar wind to be transported back toward the outer edge of the heliosphere. They are likely accelerated by the heliospheric termination shock to become ACRs,  although the precise acceleration mechanism has been a topic of debate, especially after recent observations by Voyagers 1 and 2. The energies of ACR protons can reach  $\sim$100 MeV \citep[e.g.][]{CummingsAC_1998}. \cite{JokipiiJR_1998} already argued that these highest energies are gained from electrostatic potential while drifting along the shock surface, which they estimated as 200 -- 300 MeV. In our case, we use  $B = 0.02 - 0.1$ nT and $V = 250 - 350$ km s$^{-1}$ as they were measured just upstream of the termination shock by the Voyager spacecraft \citep{BurlagaLF_2008}. For $L$, we use the expected average distance between the Sun and the termination shock along the `nose' direction, i.e., $80 - 120$ AU, or $(1.2 - 1.8)\times 10^{13}$ m. As a result, we obtain $\varepsilon_{\rm H} = 60 - 630$ MeV.

\subsection{Solar Plasmas}\label{sec:solar}

In solar flares, 
the Bremsstrahlung emission from energetic electrons produces the X-ray/$\gamma$-ray continuum up to a few tens of MeV \citep[e.g.][]{LinRP_2002}. In such remote-sensing measurements, the precise evaluation of the maximum energy of electrons is confounded by the photons produced by the decay of pions \citep[e.g.][]{VilmerN_2003}. However, {\it in-situ} measurements reported $\sim$45 MeV electrons from solar flares \citep{EvensonP_1984, MosesD_1989}.

To estimate $\varepsilon_{\rm H}$ for solar flares, we consider spatially-large flares that fit into the standard model where magnetic reconnection in the corona plays a major role \citep[e.g.][]{MasudaS_1994, ShibataK_1995}. This is a Type 2 plasma environment.  We therefore adopt the values of $V$ and $B$ in the corona, even though a large part of flare emission originates from the chromospheric footpoints of flaring loop. For the flow speed $V$, we assume the typical Alfv\'{e}n speed, 1000 - 3000 km s$^{-1}$ \citep[e.g.][]{AschwandenM_2005}.  For the magnetic field, we assume $(5.0 - 50) \times 10^ 6$ nT, or $B = 50 - 500$ G  \citep[e.g.][]{AschwandenM_2005, ChenB_2020}. For the system size $L$, we assume $L = 10 - 30$ Mm, or $(1.0 - 3.0)\times 10^7$ m, which is half the loop-size of spatially-large flares \citep[e.g.][]{AschwandenMJ_1996, AschwandenM_2005, KruckerS_2008_review, OkaM_2015}.  Alternatively, we could use the length of flare arcade (or flare ribbon) that extends in the direction of reconnection electric field. It can be $\sim200$ Mm or even larger.  However, we keep the definition outlined in Section \ref{sec:method} and use the flare loop size. In fact, hard X-ray sources are usually isolated or fragmented and do not extend along the flare arcade \citep[e.g.][]{AsaiA_2002, ShiG_2024}, suggesting that the acceleration region do not extend as long as the length of the arcade. With these parameters, we obtain $\varepsilon_{\rm H} = 0.05 - 5$ TeV.

For high-energy protons of solar origin, we consider observations by ground-based, cosmic-ray monitors. During solar eruptive events including solar flares and coronal mass ejections (CMEs), significant enhancements of GeV proton flux are observed, which are referred to as Ground Level Enhancements (GLEs) \citep[e.g.][]{CarmichaelH_1962}. To date, protons of up to $\sim$30 GV in rigidity (approximately 30 GeV in energy) have been detected \citep{LovellJL_1998}. One intriguing aspect of these observations is that the spectra often exhibit a roll-off, indicating that the power-law spectra are unlikely to extend further, even with improved measurements of GLEs. Therefore, we do not consider the observed value of  30 GeV as a lower limit.

Unfortunately, the  origin of GLE protons remains ambiguous. They can be produced by solar flares (in which magnetic reconnection plays an important role), CME-associated shocks, or both \citep{KahlerS_1994, VashenyukEV_2011, AschwandenM_2012, GuoJingnan_2023, McCrackenKG_2023}. Some observations and modeling indicate that GLEs often exhibit a two-step evolution: a prompt enhancement possibly due to a solar flare, followed by a gradual decay possibly due to a CME shock \citep[e.g.][]{VashenyukEV_2011}. However, because of the lack of consensus on the origin, we assume that both solar flares and CMEs produce protons of $\sim$30 GeV. It should be noted that recent observations by Fermi's Large Area Telescope (LAT) reported $\gamma$-ray spectra that are consistent with the decay of pions produced by $>$ 300 MeV protons \citep{AjelloM_2021}. They found at least two distinct types of $\gamma$-ray emission: prompt-impulsive and delayed-gradual, which could be related to the two-step evolution of GLE events. 

To estimate $\varepsilon_{\rm H}$ for CMEs (and associated shocks), we take into account the statistical study by \cite{KahlerS_1994}. They reported that GLEs exhibited a peak when the height of associated CME reached $5 - 15 R_\odot$ \citep{KahlerS_1994}. Then, we use $L = 5 - 15 R_\odot$, or $(3.5-10)\times 10^9$ m, combined with $B = 300-10000$ nT, or 0.003 - 0.1 G \citep{BirdMK_1990}, and $V = 1000 - 4000$ km s$^{-1}$ \citep{SmartDF_1985}, to obtain $\varepsilon_{\rm H} = 1 - 420$ GeV for CMEs in Table \ref{tab}.

\subsection{Astrophysical Plasmas}\label{sec:astro}

\subsubsection{SN 1006}

Supernova remnants (SNRs) are one of the most important astrophysical plasma environments in the context of the Hillas limit. This is because SNRs are promising sites for cosmic ray production up to the `knee' energy ($\sim$TeV) or even beyond, and whether they can actually achieve this energy  has been a subject of intense debate, both observationally and theoretically \citep[e.g.][]{LagagePO_1983, BellAR_2004, HillasAM_2005,  DruryLOC_2012, SuzukiH_2022, DiesingR_2023, TaoM_2024}.

The Hillas limit $\varepsilon_{\rm H}$ for SN 1006 is obtained as follows, using our definitions for Type 1 environment. The remnant of SN 1006 exhibits a shell structure with the average radius of $\sim 15\arcmin\,$ \citep[e.g.][]{GardnerFF_1965, KoyamaK_1995, BambaA_2003}. Using the distance of $\sim2$ kpc \citep[e.g.][]{GreenDA_2019, WinklerP_2003} where 1 pc = $3.1 \times 10^{16}$ m, we have $L \sim 2.7\times10^{17}$m. The remnant is interpreted as a shock front expanding at the speed of 0\arcsec.28 - 0\arcsec.48 per year \citep{WinklerP_2003, KatsudaS_2009} or  $2700 - 4600$ km s$^{-1}$. For the upstream magnetic field, we use the typical interstellar magnetic field of $0.1-0.5$ nT, or $1-5 \mu$G. As a result, we obtain $\varepsilon_{\rm H} \sim 73 - 620$ TeV. 

It should be noted that a higher value of $\varepsilon_{\rm H}$ can be obtained by focusing on the local conditions at the shock front. In the diffusive shock acceleration scenario \citep[e.g.][]{BlandfordR_1987}, particles are accelerated within a finite region around the shock front. Consequently, the thickness of the shell measured in X-rays and $\gamma$-rays, approximately 1\arcmin\, \citep{BambaA_2003, LiJT_2018},  should correspond to the spatial extent of the acceleration region \citep[e.g.][]{YamazakiR_2004}. For the magnetic field $B$, recent observations have found  evidence for a highly amplified value at the shell, $B\gtrsim 20$ nT \citep{TaoM_2024}, consistent with earlier expectations \citep[e.g.][]{BerezhkoEG_2002, ParizotE_2006}. Therefore, by adopting $B \sim 20$ nT and $L\sim1\arcmin$, or $1.8 \times 10^{16}$ m, we obtain a predicted maximum energy $\varepsilon_{\rm H} \sim$ 2 PeV. This is an order of magnitude higher than our first estimate.

Let us turn to the observational estimates $\varepsilon_{\rm obs}$, using X-ray and gamma-ray remote sensing. The observed X-ray spectra exhibit a clear cutoff at a photon energy of $\sim$ 0.2 keV, which can be interpreted as synchrotron emission from electrons with the maximum energy of 10 T eV in a magnetic field of 1 nT \citep{BambaA_2008}. On the other hand, observed $\gamma$-rays can be attributed to the inverse Compton scattering by electrons off photons of comic microwave background (CMB), or proton-induced pion-decay, or a combination of both. Based on a hybrid model that takes into account both mechanisms of $\gamma$-ray production, it has been reported that the maximum energies of electrons and protons are  $\varepsilon_{\rm obs, e}\sim10$ TeV and $\varepsilon_{\rm obs, p}\sim$ 100 TeV, respectively, with a magnetic field strength of 4.5 nT \citep{AceroF_2010}.  These energies match with our estimates within an order of magnitude.

\subsubsection{RX J1713.7-3946}

RX J1713.7-3946 is another shell-forming SNR, roughly 30$\arcmin$  in diameter at the distance of $\sim$ 1 kpc \citep[e.g.][]{TsujiN_2016}, giving us $L \sim 1.4\times 10^{17}$m. The expansion speed is 3900 km s$^{-1}$ \citep{TsujiN_2016}, which we use for $V$. A key difference from SN 1006 is that it is located right on the Galactic plane where the interstellar medium is much denser and possibly non-uniform. In fact, a part of the shell is moving substantially slower, indicating an ongoing interaction with local clouds \citep{TanakaT_2020}. Furthermore, the presence of  dense interstellar material appears to enable the so-called hadronic channel, in which accelerated protons collide with ambient protons to produce pions. These pions subsequently decay into a pair of 70 MeV (in rest frame) gamma-ray photons, thereby contributing significantly to the observed flux of gamma-ray emission. For the upstream magnetic field, we assume 0.1 - 1 nT, which includes values slightly higher than those we used for SN 1006, considering the presence of denser materials. Then, we predict the maximum energy of particles as $\varepsilon_{\rm H} \sim$ 53 - 530 TeV, very similar to that of SN 1006. Alternatively, we can again consider the local condition with amplified magnetic field of $B \sim$10 nT \citep{UchiyamaY_2007} and shell thickness  $L \sim 0.5\arcmin$ \citep{TanakaT_2020} to obtain $\varepsilon_{\rm H} \sim$ 170 TeV, a value consistent with the above estimation.

For observations, the spectra have been measured with both X-rays \citep[e.g.][]{TakahashiT_2008, YuanQiang_2011, TsujiN_2019} and $\gamma$-rays \citep[e.g.][]{AbdoAA_2011_RXJ1713, YuanQiang_2011, AbdallaH_2018_RXJ1713}. The observed maximum energy of particles is obtained as  $\varepsilon_{\rm obs, p}\sim \varepsilon_{\rm obs, e}\sim$ 100 TeV, although it can vary within an order of magnitude depending on the spectral model employed, including how much fraction of ions contribute to the photon emissions ({\it hadronic} versus {\it leptonic}). In this regard, a recent study that takes into account the column density of the local interstellar medium concludes that ions and electrons constitute about 70\% and 30\% of the total $\gamma$-ray emissions, respectively \citep{FukuiY_2021}. This result provides valuable evidence for proton acceleration  in astrophysical settings.

\subsubsection{Crab Nebula}

The Crab Nebula is also a supernova remnant, but the presence of a spinning pulsar at its center makes it very different from more ordinary shell-type  SNRs \citep[e.g.][and references therein]{LongairM_2011, BuhlerR_2014}. A distinctive feature in its X-ray image is the torus and the associated inner ring \citep{WeisskopfMC_2000}, which is interpreted as the termination shock formed as  the pulsar wind --- a magnetized, relativistic flow of electrons and positrons from the pulsar --- collides with the surrounding nebula \citep{KennelCF_1984_Confinement}. To estimate $\varepsilon_{\rm H}$, we use the radius of the inner ring $\sim$ 0.14 pc, or 4.3$\times 10^{15}$ m \citep{WeisskopfMC_2000} for $L$ and the light speed $c \sim 3 \times 10^5$ km s$^{-1}$ for $V$. The magnetic field $B$ in the post-shock region can be estimated from the spectral break that appears in the synchrotron spectrum at $\nu \sim 10^{13}$ Hz. Assuming that the spectral break represents the condition of the synchrotron cooling time being comparable to the age of the nebula, we can obtain $B \sim$30 nT  \citep{MarsdenPL_1984}. Similar estimates in the range of $10 - 30$ nT have been obtained by other studies \citep[e.g.][and references therein]{BuhlerR_2014}. To further estimate the magnetic field in the upstream region of the shock, we adopt the compression ratio 7 for a relativistic, strictly perpendicular shock and obtain $B \sim$ 1 - 4 nT. Combining the parameters, we obtain the Hillas limit of $\varepsilon_{\rm H} \sim 2 - 6$ PeV.

\mcl{Regarding observations, the electromagnetic spectrum from Crab Nebula has been observed over a wide range of energies from radio to $\gamma$-ray ranges, revealing both components of synchrotron and inverse Compton emissions \citep[e.g.][]{ZhangXiao_2020, AharonianF_2024}. According to these observations,} the synchrotron component exhibits a clear cutoff at the photon energy of $\sim 50$ MeV. Using the magnetic field $30$ nT obtained above and the formula for the synchrotron critical frequency, we find that this photon energy corresponds to an electron energy of $\varepsilon_{\rm obs, e} \sim2$ PeV. The inverse Compton component extends beyond 0.1 PeV \citep{AharonianF_2024}, which corresponds to the electron energy of $\varepsilon_{\rm obs, e} \sim 0.5$ PeV. However, the cutoff is not visible in the spectra and so we regard this value as a lower-limit of $\varepsilon_{\rm obs, e}$.

\mcl{It should be noted that there may be a non-negligible contribution to the spectrum from energetic protons, although the leptonic model has been widely accepted \citep[e.g.][]{AtoyanAM_1996}. In this regard, a recent broad-band spectral analysis considered hadronic photons from the neutral pion-decay process and showed that the fraction of energy converted into energetic protons is likely $<$0.5\% although it could be as high as 7\% if only the $\gamma$-ray data are used \citep{ZhangXiao_2020}. Therefore, we consider that the the plasma in the Crab Nebula consists of predominantly electrons and positron. Consequently, we did not include protons in Table \ref{tab}.}

\subsubsection{Crab Pulsar}

The Crab Pulsar is not spatially resolved by remote-sensing measurements, and much of our evaluation of the Hillas limit is based on the arguments and parameters described in classical literature \citep[e.g.][]{GoldreichP_1969, LongairM_2011, BuhlerR_2014}. The Crab Pulsar produces intense beams of radiation through acceleration of particles within its magnetosphere. The particle acceleration mechanism remains largely unknown, and there exist  different models, such as polar cap, outer gap, and slot gap models, depending on the assumed location of the acceleration site. An important spatial scale in pulsar's magnetosphere is the radius of the light cylinder $R_{\rm LC}$. This is where the speed of the magnetic field co-rotating with the pulsar reaches the light speed, so that $R_{\rm LC} = c/\Omega \sim 1.5 \times 10^6$m where $\Omega$ = 200 rad s$^{-1}$ is the pulsar's angular velocity. Therefore, to estimate the predicted maximum energy of particles,  we adopt the definitions for Type 3 environment and use $L = R_{\rm LC}$ and $V = c \sim 3 \times 10^5$ km s$^{-1}$. For the magnetic field strength $B$, it scales like $\propto r^{-3}$ where $r$ is the distance from the pulsar  and $B \sim 4 \times 10^{17}$ nT, or $4 \times 10^{12}$ G, at the equatorial surface of the neutron star. Thus, we obtain 10$^{11}$ nT at the surface of the cylinder. Then, the predicted maximum energy becomes $\sim 45$ PeV. 

The Crab Pulsar produces pulsed emissions,  and  photon spectra that extends up to $\sim$ TeV have been measured, despite the flux being two orders of magnitude smaller than that of the nebula \citep{AnsoldiS_2016, YeungPKH_2024}. Assuming inverse Compton scattering off photons of cosmic microwave background (CMB), this maximum photon energy corresponds to the maximum particle energy of $\varepsilon_{\rm obs} \sim 15$ TeV. We consider this as a lower-limit of $\varepsilon_{\rm obs}$.

\subsubsection{Cygnus A}

As touched on in Section 2, jet-terminal lobes of radio galaxies (particularly of the so-called Fanaroff-Riley Type II, or FR-II) are also an interesting astroplasma environment. It is generally considered that particle acceleration occurs predominantly at the termination (reverse) shock, creating hot-spots in a central region of their radio lobes \citep[e.g.][]{CarilliCL_1996, CasseF_2005}. In this study, we specifically consider Cygnus A (3C405) because it is one of the brightest radio galaxies and has been studied intensively from various approaches. Based on observations by, for example, \cite{MeisenheimerK_1997} and \cite{SteenbruggeKC_2008}, combined with our guideline for the Hillas limit (Section \ref{sec:method}),  we use the estimated radius of the hot-spots, $L = 1.1 - 1.5$ kpc, or $(3.4 - 4.6) \times 10^{19}$ m, and the jet speed of $V = (0.24-0.30)c$, or $72000 - 90000$ km s$^{-1}$. For the magnetic field $B$, we assume its value in the hot-spot 40 nT \citep{MeisenheimerK_1997}, and divide it by the shock compression ratio $3.9 - 4.3$ to obtain the shock upstream value $9 - 10$ nT. These numbers bring the predicted maximum energy $\varepsilon_{\rm H}$ very high, 22 - 42 EeV, up to the range of ultra-high-energy cosmic rays (UHECRs), $\gtrsim$ EeV. In fact, radio galaxy lobes was one of the candidates for the site of UHECR production based on earlier evaluations of the Hillas limit \citep[e.g.][]{HillasAM_1984, RachenJP_1993, NormanCA_1995}. 

In contrast to the above predictions, the observed maximum energy of electrons is much lower, $\varepsilon_{\rm obs} \sim$ 30 GeV in the hot-spots of Cygnus A \citep{MeisenheimerK_1997}. Similar values of $\varepsilon_{\rm obs}$ have been obtained for the lobe regions \citep{SteenbruggeKC_2008, YajiY_2010}. This issue is not specific to Cygnus A, because FR-II radio galaxies generally  show synchrotron emissions predominantly in the radio range and do not extend to the higher-energy X-ray range. The observations in the X-ray range are interpreted by the inverse Compton scattering off CMB photons by the relativistic electrons \citep[e.g.][]{FeigelsonED_1995, KanedaH_1995}. Therefore, the observed maximum energies are robust, and it is clear that electrons do not reach the Hillas limit in the hot-spots of FR-II radio galaxies.  We  discuss this  gap in Section \ref{sec:discussion}.

\begin{figure*}
\plotone{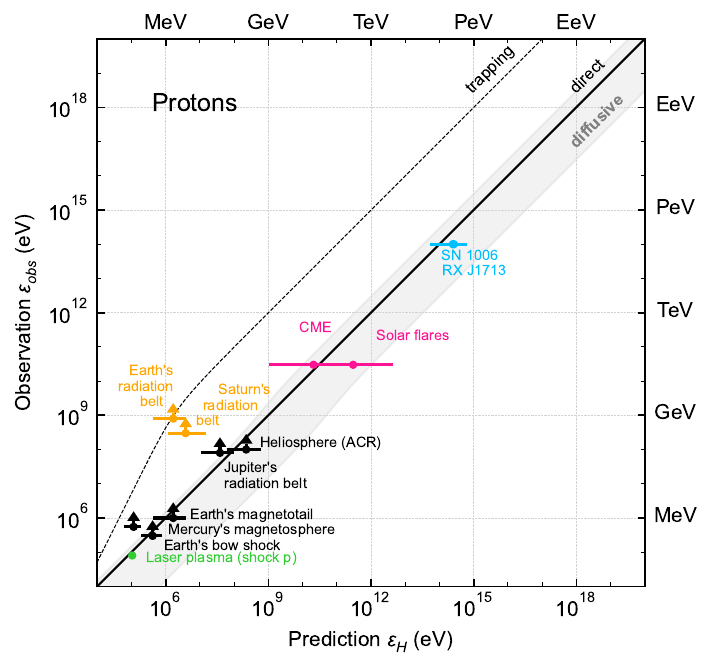}
\caption{The maximum energy of protons in various plasma environments, as compared between the Hillas limit prediction (horizontal axis) and observations (vertical axis). See Table \ref{tab} for the values used and Section \ref{sec:data} for their justification. The black line is the identity line and represents the scenario of direct, one-shot acceleration. The gray shaded region represents the scenario of stochastic and diffusive acceleration with a strong scattering, $\eta=1-100$. The dashed curve represents the trapping condition $r_g \leq L$ for the case of $\psi=c/V=1000$ or $V=300$ km s$^{-1}$.
\label{fig:Heliosphere_Protons}}
\end{figure*}
\begin{figure*}
\plotone{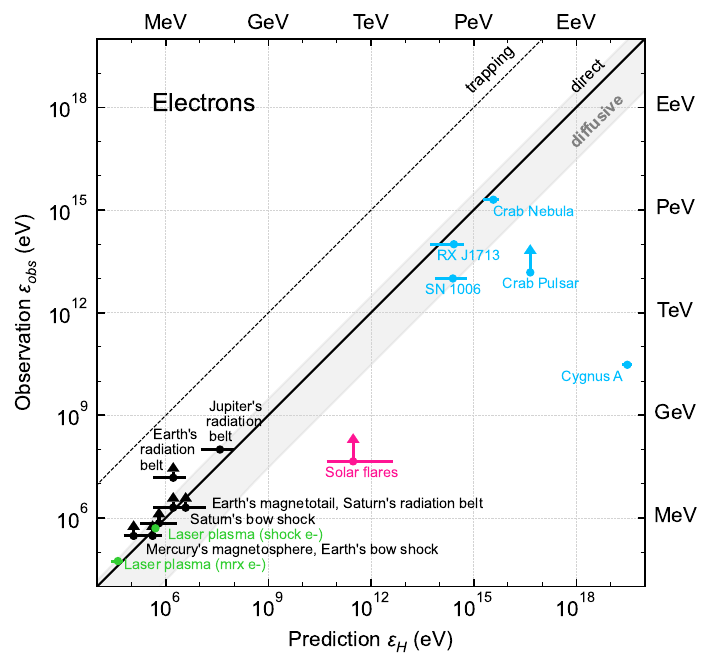}
\caption{The maximum energy of electrons in various plasma environments, with the same format as Figure \ref{fig:Heliosphere_Protons}.
\label{fig:Heliosphere_Electrons}}
\end{figure*}

\subsection{Laboratory Plasmas}\label{sec:laboratory}

Laboratory experiments provide unique opportunities to study plasma phenomena in a controlled manner. In particular, recent developments in laser facilities have enabled experimental studies of particle acceleration at shocks and reconnection in a collisionless condition. 

\cite{YaoW_2021} observed 80 keV protons at a shock front generated by a laser-driven, fast flow of plasma expanding into an ambient cloud of plasma. Using an earlier phase of the evolution when the shock front was still moving at a high speed, they evaluated the Hillas limit with the shock speed $V \sim$1500 km, the shock size $L\sim (3-4)\times 10^{-3}$ m, and the ambient magnetic field $2 \times 10^{10}$ nT, to obtain $\varepsilon_{\rm H}\sim 90 - 120$ keV which is comparable to the observed energy.

\cite{FiuzaF_2020} detected 500 keV electrons at a shock front generated by an interaction of laser-driven, counter-streaming flows of plasma. They reported the Alfv\'{e}n Mach number $M_A \sim$400. They also evaluated the Hillas limit with the following parameters: the flow speed $V \sim$1000 km s$^{-1}$, the transverse size of the shock $L \sim 5 \times 10^{-3}$ m, and the magnetic field $\sim10^{11}$ nT, or 1 MG, leading to the expected maximum energy of  $\varepsilon_{\rm H} \sim$500 keV, well consistent with the observation.

\cite{ChienA_2023} reported a detection of $40-70$ keV electrons during magnetic reconnection in a laser-driven plasma experiment. The reconnection geometry was formed by currents induced in two U-shaped coils by the same laser-driven plasma. During the initial `push' phase, when the currents were still increasing, the reconnection was more strongly driven than compared to a later phase. Their simulations of this experiment indicated that the peak rate of reconnection was $\alpha_{\rm R} \sim$0.6 where $\alpha_{\rm R}$ is defined as the inflow speed $V_{in}$ normalized by the Alfv\'{e}n speed $V_A$, i.e., $\alpha_{\rm R}=V_{in}/V_A$. Using data from  this earlier phase, along with the inferred peak value of $\alpha_{\rm R} \sim 0.6$, they calculated the reconnection electric field to be $0.6V_AB_0$, where $V_A \sim 500 - 1100$ km s$^{-1}$ and $B_0\sim 5 \times 10^{10}$ nT. With a system size of $\sim 10^{-3}$ m (or 1.4 times the ion inertia length), they derived a predicted maximum energy of 13-30 keV, which is roughly half  the observed energy. 

A caveat here is that this prediction  was based on the inflow speed of 0.6$V_A$, whereas our Hillas limit evaluation is based on the outflow speed, or equivalently $V_A$ (Section \ref{sec:method}). Thus, we estimate $\varepsilon_{\rm H}$ without the factor 0.6 and obtain $\varepsilon_{\rm H}\sim 25 - 55$ keV, which is comparable to the observed values. It should also be noted that, according to \cite{ChienA_2023}, the system size was so small that the observed reconnection was deeply in the electron-only regime, where ions are decoupled from the process and do not exhibit ion-scale jets \citep{PhanTD_2018}. In the presented systematic review, this is the smallest physical system used for testing the Hillas limit.

\section{Results}\label{sec:results}

All the values compiled in Table \ref{tab} are visualized in Figures \ref{fig:Heliosphere_Protons} and \ref{fig:Heliosphere_Electrons} for protons and electrons, respectively. In both panels, the horizontal axis shows the maximum energy $\varepsilon_{\rm H}$ predicted by Eq.(\ref{eq:hillas-V}), derived from the parameters $V$, $B$, and $L$, whereas the vertical axis shows the observed maximum energy $\varepsilon_{\rm obs}$. For those plasma environment in which $\varepsilon_{\rm obs}$ is the lower-limit of the observations, we added an upward arrow. The black solid line is the identify line of $\varepsilon_{\rm obs} = \varepsilon_{\rm H}$. It is labeled `direct' because the Hillas limit can be interpreted by `direct' (one-shot) acceleration.

\begin{figure}
\epsscale{1.2} 
\plotone{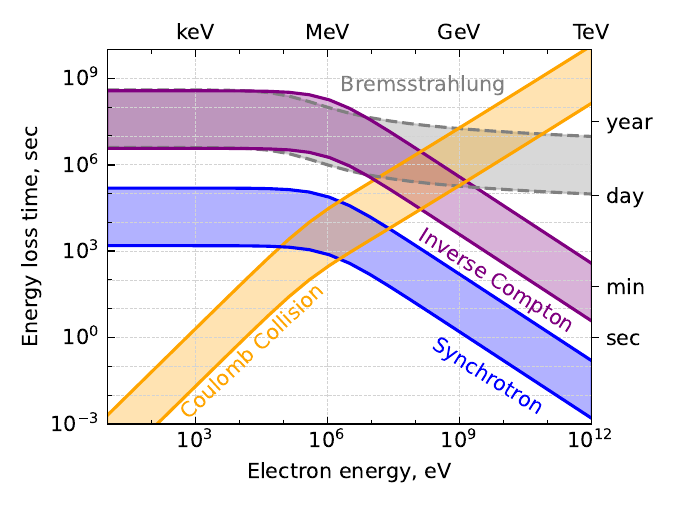}
\caption{Energy loss time for fast electrons in the solar corona, demonstrating the importance of synchrotron cooling in the relativistic regime. Four different processes (as annotated) are considered with  densities  $n=10^{8} - 10^{10}$cm$^{-3}$ and magnetic fields $B=(5 - 50) \times 10^{6}$ nT, or $50 - 500$ G . In the case of inverse Compton, we considered scattering off photons of solar origin instead of the Cosmic Microwave Background (CMB).
\label{fig:timescale}}
\end{figure}

The gray shaded area indicates `diffusive' acceleration, as approximated by Eq.(\ref{eq:terasawa}) and more accurately expressed as follows.  The spatial diffusion coefficient is $D = V\lambda = \eta p v/3qB$, where $p$ is the particle momentum.  The particle kinetic energy is  $\varepsilon_{\rm obs} = \sqrt{p^2c^2 + m^2c^4} - mc^2$. Then, from the condition of $\lambda \leq L$ where $\lambda$ is the spatial scale of the acceleration region, combined with $\varepsilon_{\rm H} = qVBL$, we eventually obtain the  maximum energy from the case of $\lambda = L$ as
\begin{equation}
    \varepsilon_{\rm obs} = \varepsilon_{\rm H}^{\prime} + \sqrt{\varepsilon_{\rm H}^{\prime2} + m^2c^4} - mc^2
\end{equation}
where $\varepsilon_{\rm H}^{\prime} = (3/2\eta) \varepsilon_{\rm H}$. The shaded region in Figures \ref{fig:Heliosphere_Protons} and \ref{fig:Heliosphere_Electrons} shows this condition with the range of $\eta = 1 - 100$. In the limit of $\varepsilon_{\rm H} \gg mc^2$, this condition reduces to Eq.(\ref{eq:terasawa}) or $\varepsilon_{\rm obs} = (3/\eta) \varepsilon_{\rm H}$. 

The dashed line indicates the `trapping' condition or the {\it hard} limit, as approximated by Eq.(\ref{eq:hillas-c}) and more accurately expressed as follows. From the condition of $r_g \leq L$ where $r_g = p/qB$ is the gyro-radius, we eventually obtain
\begin{equation}\label{eq:hillas-c-full}
    \varepsilon_{\rm obs} = \sqrt{\psi^2 \varepsilon_{\rm H}^2 + m^2c^4} - mc^2
\end{equation}
where $\psi = c/V$. The dashed curve in Figures \ref{fig:Heliosphere_Protons} and \ref{fig:Heliosphere_Electrons} shows this condition with $\psi = 1000$. This corresponds to $V$=300 km s$^{-1}$, which is close to the lower limit of the typical range of $V$ in many plasma environment (See Table \ref{tab}). In the limit of $\varepsilon_{\rm H} \gg mc^2$, Eq.(\ref{eq:hillas-c-full}) reduces to Eq.(\ref{eq:hillas-c}) or $\varepsilon_{\rm obs} = \psi \varepsilon_{\rm H} = \varepsilon_{\rm limit}$. 

It is evident from Figure \ref{fig:Heliosphere_Protons} that the Hillas prediction generally agrees with the observed maximum energies of protons within an order of magnitude. An exception is the radiation belts at Earth and Saturn (orange), as their maximum energies substantially exceeds the Hillas limit (solid line). We consider this excess is due to externally-supplied CRAND protons. Their observed energies are still lower than or comparable to the hard limit (dashed line). 

The Hillas prediction also agrees with the observed maximum energies of electrons within an order of magnitude, as shown in Figure \ref{fig:Heliosphere_Electrons}, although there is a few large deviations. In Cygnus A, the observed maximum energy $\varepsilon_{\rm obs, e}$ is nearly 9 orders of magnitude lower than the prediction.  As described in Section \ref{sec:astro}, we do not consider this a lower limit. For solar flares and the Crab Pulsar, the deviation is roughly $\sim$4 orders of magnitude, but the observed values can be regarded as a lower limit. Therefore, it is still possible that the particle energies do reach the Hillas limit in these environments.

To further evaluate the validity of the Hillas limit, we performed a multivariate analysis where the dependencies of the observed maximum energy $\varepsilon_{\rm obs}$ on the parameters $V$, $B$, and $L$ are modeled as 
\begin{equation}
    \varepsilon_{\rm obs} = DV^aB^bL^c
\end{equation}
where $D$ is a constant and the indices $a$, $b$, and $c$ are free parameters to represent those dependencies. This is, in a sense, a purely empirical, bottom-up analysis with no prior knowledge of particle acceleration theories. Although some values of $\varepsilon_{\rm obs}$ are considered as lower limits, we took the logarithm of the above expression and  performed a linear regression, treating all values equally\mcl{, after removing obvious outliers:  Earth's radiation belt and Saturn's radiation belt for protons and solar flares, the Crab Pulsar, and Cygnus A for electrons}. The resultant best-fit model for protons is
\begin{equation}
    \varepsilon_{\rm obs, p} = (9.6\times 10^{-4})\, V^{1.47\pm0.56} B^{0.79\pm0.10} L^{0.86\pm0.07}
\end{equation}
It is evident that all indices are close to unity, indicating that the observed maximum energies of particle depend almost linearly  on $V$, $B$, and $L$, consistent with Eq. (\ref{eq:hillas-V}). For electrons, we performed the same analysis and obtained
\begin{equation}
    \varepsilon_{\rm obs, e} = (2.6\times 10^{-2}) \, V^{1.27\pm0.23} B^{0.78\pm0.09} L^{0.82\pm0.06}
\end{equation}
Again, the indices are close to unity.

\mcl{Therefore, we} conclude that the Hillas Limit actually holds in many particle acceleration environments, even when the parameters $B$ and $L$ vary greatly  \mcl{across  many orders of magnitude. However, we emphasize again that we treated all values equally, including those considered  lower limits. Thus, these multivariate analysis results should be validated in the future as improved measurements of the maximum energy of particles become available. Additionally, we excluded known outliers that cannot be explained by the Hillas limit alone. For such outliers, additional physical processes must be considered, such as an external supply for protons and energy loss and weak scattering conditions for electrons.  These topics are discussed in the next section.}

\section{Discussion \label{sec:discussion}}

While it has been established that accelerated non-thermal particles often exhibit a power-law spectrum, how far the power-law spectrum extends and what limits the particle energy have remained unclear. Here, in this study, we have confirmed earlier suggestions \citep{MakishimaK_1999, TerasawaT_2001} that the highest observed energy of particles $\varepsilon_{\rm obs}$ does reach the Hillas limit $\varepsilon_{\rm H}$.  In fact, they often match quite well, i.e., $\varepsilon_{\rm obs} \sim \varepsilon_{\rm H}$, over a very wide range of parameters, despite the fact that many of the observed values should be regarded as lower limits. The multivariate analysis that treat all $\varepsilon_{\rm obs}$ values equally returned indices that are consistent with the Hillas limit expressed as Eq.(\ref{eq:hillas-V}).  This universal scaling law becomes even more impressive when we recall the earlier argument that the Hillas limit, which initially appears as a one-shot direction acceleration by a system-wide electric field (Eq.(\ref{eq:hillas-V})), can actually be interpreted within the scenario of stochastic and diffusive acceleration with strong scattering \citep{TerasawaT_2001}, as expressed in Eq.(\ref{eq:terasawa}). Furthermore, the {\it hard} limit based on the condition of $r_g \leq L$ also appears to nicely explain a few cases of planetary radiation belt that clearly exceeded the Hillas limit. In this section, we first explorer the possible implications of this scaling law on the underlying mechanism of particle acceleration. We then discuss some of the deviations from the Hillas limit that we have identified in our systematic review of the maximum energy of particles.

\subsection{Particle Acceleration Mechanisms}

We emphasize that it is possible for a stochastic acceleration to work simultaneously with energization by the system-wide motional electric field. This is perhaps most clearly described by \cite{JokipiiJR_1987} who showed theoretically that, if the cross-field diffusion is sufficiently strong across a quasi-perpendicular shock,  particles can be energized as they drift along the $\mathbf{V} \times \mathbf{B}$ electric field while experiencing diffusive shock acceleration, resulting in a higher rate of energy gain than in the case of a quasi-parallel shock. 

It has also been suggested that both coherent and stochastic components of acceleration work for particle acceleration in MHD turbulence due to magnetic reconnection \citep[e.g.][]{AmbrosianoJ_1988, DmitrukP_2003}. In their work, a normalized version of Eq.(\ref{eq:hillas-V}) is used, following \cite{MatthaeusWH_1984}, and the system size $L$ is replaced with the Alfv\'{e}n transit time $\tau_A$. Then, it was demonstrated with test particle simulations  that,  as $\tau_A$ increases, the maximum energy of particles also increases in MHD turbulence \citep{DmitrukP_2003}. 


Similarly, two-dimensional (2D), full particle simulations  have shown that  the highest energy of particles is matched with the prediction of $qE_R\Delta Z$ where $q$ is the particle charge, $E_R$ is the reconnection electric field, and  $\Delta Z$ is the displacement of particles in the direction of the reconnection electric field (See Figure 11 of \cite{OkaM_2010_coa}). The match was observed in simulations of multi-island coalescence where particles change their direction of motion (and thus experience scattering) at the merging point. More formal treatment of particle acceleration in quasi-2D interacting magnetic islands can be found in,  for example,  \cite{ZankGP_2014}. A key point here is that the same effect might take place in 3D magnetic reconnection with turbulence. Therefore, we speculate that, even in a turbulent and diffusive environment, a small number of fortuitous particles can still experience energization while moving across the system as implied by the Hillas limit.

\subsection{Electrons in Radio Galaxy Lobes}

While it is difficult to distinguish the two different acceleration mechanisms with our dataset alone, more detailed investigations of radio galaxy lobes (including the hot spots) might provide an important clue. This is because there is a strikingly large discrepancy between $\varepsilon_{\rm obs}$ and $\varepsilon_{\rm H}$ for electrons in Cygnus A.  The gap is roughly 9 orders of magnitude. Such a large gap  exists in other FR-II radio galaxies as well, and it has been an important subject of research  \citep[e.g.][and references therein]{CasseF_2005, AraudoAT_2018}. It was argued that the observed maximum energy cannot be explained by synchrotron cooling even after considering magnetic field amplification at the shock front and that the hot-spots of FR-II radio galaxies are indeed poor accelerators of particles \citep{AraudoAT_2016, AraudoAT_2018}. 

This is in stark contrast to the Heliosphere (ACR) and the Crab Nebula where particles are accelerated at a reverse shock, similar to hot-spots of radio galaxies, and yet the maximum energies of particles reach very close to the Hillas limit. At and around the termination shock of the Heliosphere and the Crab nebula, magnetic reconnection could occur frequently due to the intrinsic presence of current sheets in the solar and pulsar winds. Such current sheets, when compressed by the termination shock, can lead to enhanced turbulence via magnetic reconnection and associated particle acceleration \citep[e.g.][]{DrakeJF_2010, ZankGP_2014, ZankGP_2015, LuYingchao_2021}. On the other hand, jets in radio galaxies do not necessarily carry current sheets, although the precise magnetic field structure within the jets is unknown. We therefore conjecture that magnetic reconnection is lacking at the termination shock of radio-galaxy hot-spots, thereby reducing the chance of particle scattering by reconnection-induced turbulence. 

Further investigations of magnetic field structure in the hot-spots may lead to a better understanding of the gap between observations and the Hillas limit of the maximum energy of particles of radio lobes,  and more generally, to a deeper insight into the particle acceleration mechanisms in various plasma environments.

\subsection{Electrons in Solar Flares}

The maximum energy of electrons in solar flares is also much smaller than what is predicted by the Hillas limit. This is particularly important because protons do reach the energy much closer to the Hillas limit prediction. It should be emphasized again that the highest energy electrons of solar origin were observed by {\it in-situ} measurements that are limited by the sensitivity and energy coverage of the instrument. In general, measuring high energy electrons in space is more difficult than measuring protons, due to various reasons including the effect of background noise (secondary electrons) caused by high-energy protons. Therefore, it is important to continue improving the diagnostics of high-energy electrons of solar origin. If we determine there really is a large discrepancy between the Hillas limit and the observed maximum energy of electrons, it indicates that a physical process that is unique to electrons, either acceleration or energy loss, is at play. Because electrons reach the Hillas limit energy levels in many space plasma environments and there is no reason to believe turbulence is weaker in solar plasmas than in space plasmas, we tentatively consider the possibility that the maximum energy of solar electrons is limited by radiative energy losses. 

Figure \ref{fig:timescale} shows the timescale of different energy-loss processes, as a function of electron energy, obtained by using formulas summarized by \cite{LongairM_2011}. For Coulomb collisions, we used a formula provided by \cite{SaintHilaireP_2005}. It is evident that, while collisions dominate in the non-relativistic regime, energy loss by synchrotron emission dominates in the relativistic regime \citep[See also Appendix of][]{KruckerS_2008_review}. 

\mcl{In fact,} the timescale of synchrotron energy loss for a particle of energy $\varepsilon$ is expressed as
\begin{equation}
    \tau_{\rm synch} \sim 38 \left(\frac{B}{100\,{\rm G}}\right)^{-2} \left(\frac{\varepsilon}{\rm GeV}\right) ^{-1}\,{\rm sec}
\end{equation}
This is shorter than the typical duration of large-scale impulsive flares ($\sim$ minutes) but longer than that of X-ray pulsations ($\sim$0.5s) which are thought to be individual sequences of energy-release \citep[e.g.][]{AschwandenM_2005}. 

\mcl{This can also be compared to the energy loss time due to inverse Compton scattering off photons of solar origin. With the solar luminosity of $L_{\odot} = 3.83\times10^{33}\, {\rm erg\, s^{-1}}$, its energy density at the solar surface is $L_{\odot}/4\pi R_{\odot}^2c = 1.31 \times 10^{18} {\rm eV\, m^{-3}}$. Then, the energy loss time is
\begin{equation}
    \tau_{\scriptscriptstyle \rm IC, \odot} \sim 7\times10^3\left(\frac{\varepsilon}{\rm GeV}\right) ^{-1}\,{\rm sec}
\end{equation}
Similarly, if we consider inverse Compton scattering off CMB photons with an energy density of $2.65\times 10^5 {\rm eV\,m^{-3}}$, the energy loss time is 
\begin{equation}
    \tau_{\scriptscriptstyle \rm IC, CMB} \sim 4\times10^{16}\left(\frac{\varepsilon}{\rm GeV}\right) ^{-1}\,{\rm sec}
\end{equation}
These timescales are much larger than that of energy loss by synchrotron emission (see also Figure \ref{fig:timescale}), and electron energies could reach the Hillas limit if they were to lose energy only by the inverse Compton emission. Therefore, synchrotron emission remains as the possible mechanism of radiative energy loss during electron acceleration in the solar corona. }

Now, if we consider the one-shot direct acceleration \mcl{by the electric field $E\sim VB$ in the corona, where $V\sim 3000$ km/s and $B\sim$ 100 G, we can readily see that electrons would reach relativistic energies and travel over the distance of $L \sim$ 20 Mm almost instantly within  $\sim9\times 10^{-5}$s. Therefore, the energy loss by synchrotron is negligible in the direct acceleration scheme.}

\mcl{This is not necessarily the case for} the diffusive acceleration scheme from which Eq.(\ref{eq:terasawa}) is derived. \mcl{The acceleration timescale is $\sim D/V^2$, and the condition $D/V^2 \lesssim \tau_{\rm synch}$ gives us}
\begin{equation}\label{eq:ultra-high-energy2}
    \varepsilon_{\rm{diffusive}}  \lesssim \frac{100}{\sqrt{\eta}}\left(\frac{B}{100\,{\rm G}}\right)^{-\frac{1}{2}} \left(\frac{V}{3000\,{\rm km/s}}\right)\, {\rm GeV}
\end{equation}
\mcl{This indicates that the electron energy will still have} no problem reaching the Hillas limit, $0.05 - 5$ TeV, depending on the parameters $B$ and $V$, in the strong scattering limit of $\eta\sim1$. However, in a pessimistic case of $B \sim$500 G, $V \sim$ 1000 km s$^{-1}$, and $\eta \sim 10^4$, the electron energy might be limited to $\sim$150 MeV, only a few times higher than the highest energy currently observed, 45 MeV.

In summary, for solar flare electrons, the discrepancy between the observed maximum energy and the Hillas limit can be explained by the synchrotron energy loss, but the synchrotron limit depends on the parameters $V$, $B$, and $L$. In particular,  the parameter values of $V$ and $B$ can vary depending on the observation methodology and the height of plasma from the solar surface. Based on Eq.(\ref{eq:ultra-high-energy2}), there is still a good chance of detecting ultra-high-energy (up to $\gtrsim$100 GeV) solar flare electrons that have not yet been detected. We call for further efforts to achieve better {\it in-situ} measurements of high-energy, solar flare electrons with higher sensitivity and wider energy coverage. We hope that future measurements can more clearly identify  the higher-energy cutoff in the energy spectra, just like the roll-off energies seen in proton spectra during GLEs. 

\subsection{Protons in Planetary Radiation Belts}\label{sec:CRAND}

In contrast to the above cases of discrepancy where electron energies fall well below the Hillas limit, proton energies that substantially exceed the Hillas limit have been observed in the radiation belts of Earth and Saturn (Figure \ref{fig:Heliosphere_Protons}). As mentioned earlier, these protons can be explained by CRAND \citep[e.g.][]{SingerSF_1958a, SingerSF_1958b, HessWN_1959, LiYuXuan_2023}. The GCR-induced neutrons produce protons, electrons, and antineutrinos through $\beta$ decay, but protons carry the largest energy and are trapped in the radiation belts. Interestingly, proton energies do not exceed the Hillas limit as much in Jupiter's radiation belt. In fact, it has already been argued that CRAND is weak in Jupiter's radiation belt \citep{KollmannP_2017}. In Jupiter's plasma environment, particles are continuously supplied from the volcanic moon Io  \citep[e.g.][and references therein]{JackmanCM_2014, GuoRuilong_2024}. Heavier ions rather than protons are the main constituent, while CRAND only provides protons and electrons. Therefore, it is not surprising that we have not seen enhanced flux of highest energy protons beyond the Hillas limit.


\section{Conclusion}\label{sec:conclusion}

We tested the Hillas limit using recent results of observations of various plasma environments, combined with renewed, consistent definitions of the key parameters $V$, $B$, and $L$. We found that particles do reach the Hillas limit in many cases, confirming earlier findings \citep{MakishimaK_1999, TerasawaT_2001}. However, there were some exceptions. Accelerated electrons in the FR-II radio galaxy Cygnus A are 9 orders of magnitude less energetic compared to the Hillas limit predictions. We speculate that this is due to possible absence of magnetic reconnection in the hot-spots and that scattering is too weak to achieve acceleration up to the Hillas limit.  Similarly, in solar flares, electrons have been observed with energies up to tens of MeV, which is well below the energy predicted by the Hillas limit. We suggest that electrons with much higher energies could be detected with improved instrumentation, although there is a possibility that the energy is limited by  synchrotron cooling. On the other hand,  the maximum energy of protons in planetary radiation belts clearly exceeds the Hillas limit due to externally supplied CRAND protons. 

It should be noted that \cite{ChienA_2023} already argued that there is a large gap between observations and the Hillas limit in many different plasma environments, where magnetic reconnection may play a major role (Type 2). In this study, we focused only on plasma environments from which we could confidently derive key parameters $V$, $B$, and $L$, as well as the observed maximum energy $\varepsilon_{\rm obs}$. Consequently, our list contained many plasma environments where shocks are likely playing a major role in energy conversion (Type 1 in Figure \ref{fig:illustration} and Table \ref{tab}). A more comprehensive test of the Hillas limit with a larger number of Type 2 plasma environments is left for future studies, and we anticipate more interdisciplinary discussions of the maximum energy of particles that use both theoretical and observational approaches.

\begin{acknowledgments}
We benefited from discussions with many experts from various fields. We particularly thank \mcl{Hantao Ji for valuable discussions in the early stage of this study,} Lyndsay Fletcher for drawing our attention to synchrotron cooling in solar plasma, and  Shunsaku Nagasawa for his help in estimating the timescales of radiative losses. We also thank Drew L. Turner for providing relevant references on radiation belts, Christopher C. Chaston and Sol\'{e}ne Lejosne for their helpful discussions on the same topic, and Katsuaki Asano for his valuable input on the Crab Nebula flares. MO was supported by NASA grants 80NSSC18K1002 and 80NSSC22K0520 at UC Berkeley. 
\end{acknowledgments}


\begin{thebibliography}{}
\expandafter\ifx\csname natexlab\endcsname\relax\def\natexlab#1{#1}\fi
\providecommand{\url}[1]{\href{#1}{#1}}
\providecommand{\dodoi}[1]{doi:~\href{http://doi.org/#1}{\nolinkurl{#1}}}
\providecommand{\doeprint}[1]{\href{http://ascl.net/#1}{\nolinkurl{http://ascl.net/#1}}}
\providecommand{\doarXiv}[1]{\href{https://arxiv.org/abs/#1}{\nolinkurl{https://arxiv.org/abs/#1}}}

\bibitem[{Abdalla {et~al.}(2018)Abdalla, Abramowski, Aharonian, Ait~Benkhali, Akhperjanian, Andersson, Angüner, Arrieta, Aubert, Backes, Balzer, Barnard, Becherini, Becker~Tjus, Berge, Bernhard, Bernlöhr, Blackwell, Böttcher, Boisson, Bolmont, Bordas, Bregeon, Brun, Brun, Bryan, Bulik, Capasso, Carr, Casanova, Cerruti, Chakraborty, Chalme-Calvet, Chaves, Chen, Chevalier, Chrétien, Colafrancesco, Cologna, Condon, Conrad, Cui, Davids, Decock, Degrange, Deil, Devin, deWilt, Dirson, Djannati-Ataï, Domainko, Donath, Drury, Dubus, Dutson, Dyks, Edwards, Egberts, Eger, Ernenwein, Eschbach, Farnier, Fegan, Fernandes, Fiasson, Fontaine, Förster, Fukuyama, Funk, Füßling, Gabici, Gajdus, Gallant, Garrigoux, Giavitto, Giebels, Glicenstein, Gottschall, Goyal, Grondin, Hadasch, Hahn, Haupt, Hawkes, Heinzelmann, Henri, Hermann, Hervet, Hinton, Hofmann, Hoischen, Holler, Horns, Ivascenko, Jacholkowska, Jamrozy, Janiak, Jankowsky, Jankowsky, Jingo, {et~al.}}]{AbdallaH_2018_RXJ1713}
Abdalla, H., Abramowski, A., Aharonian, F., {et~al.} 2018, Astronomy \& Astrophysics, 612, \dodoi{10.1051/0004-6361/201629790}

\bibitem[{Abdo {et~al.}(2011)Abdo, Ackermann, Ajello, Allafort, Baldini, Ballet, Barbiellini, Baring, Bastieri, Bellazzini, Berenji, Blandford, Bloom, Bonamente, Borgland, Bouvier, Brandt, Bregeon, Brigida, Bruel, Buehler, Buson, Caliandro, Cameron, Caraveo, Casandjian, Cecchi, Chaty, Chekhtman, Cheung, Chiang, Cillis, Ciprini, Claus, Cohen-Tanugi, Conrad, Corbel, Cutini, de~Angelis, de~Palma, Dermer, Digel, do~Couto~e Silva, Drell, Drlica-Wagner, Dubois, Dumora, Favuzzi, Ferrara, Fortin, Frailis, Fukazawa, Fukui, Funk, Fusco, Gargano, Gasparrini, Gehrels, Germani, Giglietto, Giordano, Giroletti, Glanzman, Godfrey, Grenier, Grondin, Guiriec, Hadasch, Hanabata, Harding, Hayashida, Hayashi, Hays, Horan, Jackson, Jóhannesson, Johnson, Kamae, Katagiri, Kataoka, Kerr, Knödlseder, Kuss, Lande, Latronico, Lee, Lemoine-Goumard, Longo, Loparco, Lovellette, Lubrano, Madejski, Makeev, Mazziotta, McEnery, Michelson, Mignani, Mitthumsiri, Mizuno, Moiseev, {et~al.}}]{AbdoAA_2011_RXJ1713}
Abdo, A.~A., Ackermann, M., Ajello, M., {et~al.} 2011, The Astrophysical Journal, 734, \dodoi{10.1088/0004-637x/734/1/28}

\bibitem[{Acero {et~al.}(2010)Acero, Aharonian, Akhperjanian, Anton, Barres~de Almeida, Bazer-Bachi, Becherini, Behera, Beilicke, Bernlöhr, Bochow, Boisson, Bolmont, Borrel, Brucker, Brun, Brun, Bühler, Bulik, Büsching, Boutelier, Chadwick, Charbonnier, Chaves, Cheesebrough, Conrad, Chounet, Clapson, Coignet, Dalton, Daniel, Davids, Degrange, Deil, Dickinson, Djannati-Ataï, Domainko, Drury, Dubois, Dubus, Dyks, Dyrda, Egberts, Eger, Espigat, Fallon, Farnier, Fegan, Feinstein, Fiasson, Förster, Fontaine, Füßling, Gabici, Gallant, Gérard, Gerbig, Giebels, Glicenstein, Glück, Goret, Göring, Hauser, Hauser, Heinz, Heinzelmann, Henri, Hermann, Hinton, Hoffmann, Hofmann, Hofverberg, Holleran, Hoppe, Horns, Jacholkowska, de~Jager, Jahn, Jung, Katarzyński, Katz, Kaufmann, Kerschhaggl, Khangulyan, Khélifi, Keogh, Klochkov, Kluźniak, Kneiske, Komin, Kosack, Kossakowski, Lamanna, Lemoine-Goumard, Lenain, Lohse, Marandon, Marcowith, Masbou, Maurin, {et~al.}}]{AceroF_2010}
Acero, F., Aharonian, F., Akhperjanian, A.~G., {et~al.} 2010, Astronomy \& Astrophysics, 516, \dodoi{10.1051/0004-6361/200913916}

\bibitem[{Aharonian {et~al.}(2024)Aharonian, Ait~Benkhali, Aschersleben, Ashkar, Backes, Baktash, Barbosa~Martins, Batzofin, Becherini, Berge, Bernlöhr, Bi, Böttcher, Boisson, Bolmont, de~Bony~de Lavergne, Borowska, Bradascio, Breuhaus, Brose, Brown, Brun, Bruno, Bulik, Burger-Scheidlin, Bylund, Caroff, Casanova, Cecil, Celic, Cerruti, Chambery, Chand, Chandra, Chen, Chibueze, Chibueze, Cotter, Cristofari, Devin, Djannati-Ataï, Djuvsland, Dmytriiev, Einecke, Ernenwein, Fegan, Feijen, Filipović, Fontaine, Füßling, Funk, Gabici, Gallant, Giavitto, Glawion, Glicenstein, Glombitza, Goswami, Grolleron, Grondin, Haerer, Hinton, Hofmann, Holch, Holler, Horns, Jamrozy, Jankowsky, Joshi, Kasai, Katarzyński, Khatoon, Khélifi, Kluźniak, Komin, Kosack, Kostunin, Kundu, Lang, Le~Stum, Leitl, Lemière, Lemoine-Goumard, Lenain, Leuschner, Luashvili, Mackey, Malyshev, Malyshev, Marandon, Marinos, Martí-Devesa, Marx, Mehta, Meyer, Mitchell, Moderski, Mohrmann, Montanari, Moulin, {et~al.}}]{AharonianF_2024}
Aharonian, F., Ait~Benkhali, F., Aschersleben, J., {et~al.} 2024, Astronomy \& Astrophysics, 686, \dodoi{10.1051/0004-6361/202348651}

\bibitem[{Ajello {et~al.}(2021)Ajello, Baldini, Bastieri, Bellazzini, Berretta, Bissaldi, Blandford, Bonino, Bruel, Buson, Cameron, Caputo, Cavazzuti, Cheung, Chiaro, Costantin, Cutini, D’Ammando, De~Palma, Desiante, Di~Lalla, Di~Venere, Dirirsa, Fegan, Fukazawa, Funk, Fusco, Gargano, Gasparrini, Giordano, Giroletti, Green, Guiriec, Hays, Hewitt, Horan, Jóhannesson, Kovac’Evic’, Kuss, Larsson, Latronico, Li, Longo, Lovellette, Lubrano, Maldera, Manfreda, Martí-Devesa, Mazziotta, Mereu, Michelson, Mizuno, Monzani, Morselli, Moskalenko, Negro, Omodei, Orienti, Orlando, Paneque, Pei, Persic, Pesce-Rollins, Petrosian, Piron, Porter, Principe, Racusin, Rainò, Rando, Rani, Razzano, Razzaque, Reimer, Reimer, Serini, Sgrò, Siskind, Spandre, Spinelli, Tak, Troja, Valverde, Wood, \& Zaharijas}]{AjelloM_2021}
Ajello, M., Baldini, L., Bastieri, D., {et~al.} 2021, The Astrophysical Journal Supplement Series, 252, 13, \dodoi{10.3847/1538-4365/abd32e}

\bibitem[{Alves~Batista {et~al.}(2019)Alves~Batista, Biteau, Bustamante, Dolag, Engel, Fang, Kampert, Kostunin, Mostafa, Murase, Oikonomou, Olinto, Panasyuk, Sigl, Taylor, \& Unger}]{AlvesBatistaR_2019}
Alves~Batista, R., Biteau, J., Bustamante, M., {et~al.} 2019, Frontiers in Astronomy and Space Sciences, 6, \dodoi{10.3389/fspas.2019.00023}

\bibitem[{Amano {et~al.}(2020)Amano, Katou, Kitamura, Oka, Matsumoto, Hoshino, Saito, Yokota, Giles, Paterson, Russell, Le~Contel, Ergun, Lindqvist, Turner, Fennell, \& Blake}]{AmanoT_2020}
Amano, T., Katou, T., Kitamura, N., {et~al.} 2020, Phys Rev Lett, 124, 065101, \dodoi{10.1103/PhysRevLett.124.065101}

\bibitem[{Ambrosiano {et~al.}(1988)Ambrosiano, Matthaeus, Goldstein, \& Plante}]{AmbrosianoJ_1988}
Ambrosiano, J., Matthaeus, W.~H., Goldstein, M.~L., \& Plante, D. 1988, Journal of Geophysical Research: Space Physics, 93, 14383, \dodoi{10.1029/JA093iA12p14383}

\bibitem[{Angelopoulos {et~al.}(1992)Angelopoulos, Baumjohann, Kennel, Coroniti, Kivelson, Pellat, Walker, Luhr, \& Paschmann}]{AngelopoulosV_1992}
Angelopoulos, V., Baumjohann, W., Kennel, C.~F., {et~al.} 1992, J Geophys Res-Space, 97, 4027, \dodoi{Doi 10.1029/91ja02701}

\bibitem[{Angelopoulos {et~al.}(1996)Angelopoulos, Coroniti, Kennel, Kivelson, Walker, Russell, McPherron, Sanchez, Meng, Baumjohann, Reeves, Belian, Sato, Friis-Christensen, Sutcliffe, Yumoto, \& Harris}]{AngelopoulosV_1996}
Angelopoulos, V., Coroniti, F.~V., Kennel, C.~F., {et~al.} 1996, Journal of Geophysical Research: Space Physics, 101, 4967, \dodoi{10.1029/95JA02722}

\bibitem[{Ansoldi {et~al.}(2016)Ansoldi, Antonelli, Antoranz, Babic, Bangale, Barres~de Almeida, Barrio, Becerra~González, Bednarek, Bernardini, Biasuzzi, Biland, Blanch, Bonnefoy, Bonnoli, Borracci, Bretz, Carmona, Carosi, Colin, Colombo, Contreras, Cortina, Covino, Da~Vela, Dazzi, De~Angelis, De~Caneva, De~Lotto, de~Oña~Wilhelmi, Delgado~Mendez, Di~Pierro, Dominis~Prester, Dorner, Doro, Einecke, Eisenacher~Glawion, Elsaesser, Fernández-Barral, Fidalgo, Fonseca, Font, Frantzen, Fruck, Galindo, García~López, Garczarczyk, Garrido~Terrats, Gaug, Godinović, González~Muñoz, Gozzini, Hanabata, Hayashida, Herrera, Hirotani, Hose, Hrupec, Hughes, Idec, Kellermann, Knoetig, Kodani, Konno, Krause, Kubo, Kushida, La~Barbera, Lelas, Lewandowska, Lindfors, Lombardi, Longo, López, López-Coto, López-Oramas, Lorenz, Makariev, Mallot, Maneva, Mannheim, Maraschi, Marcote, Mariotti, Martínez, Mazin, Menzel, Miranda, Mirzoyan, Moralejo, Munar-Adrover, Nakajima, Neustroev, Niedzwiecki, Nevas~Rosillo, Nilsson,
  Nishijima, Noda, Orito, Overkemping, {et~al.}}]{AnsoldiS_2016}
Ansoldi, S., Antonelli, L.~A., Antoranz, P., {et~al.} 2016, Astronomy \& Astrophysics, 585, \dodoi{10.1051/0004-6361/201526853}

\bibitem[{Araudo {et~al.}(2018)Araudo, Bell, Blundell, \& Matthews}]{AraudoAT_2018}
Araudo, A.~T., Bell, A.~R., Blundell, K.~M., \& Matthews, J.~H. 2018, Monthly Notices of the Royal Astronomical Society, 473, 3500, \dodoi{10.1093/mnras/stx2552}

\bibitem[{Araudo {et~al.}(2016)Araudo, Bell, Crilly, \& Blundell}]{AraudoAT_2016}
Araudo, A.~T., Bell, A.~R., Crilly, A., \& Blundell, K.~M. 2016, Monthly Notices of the Royal Astronomical Society, 460, 3554, \dodoi{10.1093/mnras/stw1204}

\bibitem[{Artemyev {et~al.}(2024)Artemyev, Zhang, Demekhov, Meng, Angelopoulos, \& Fedorenko}]{ArtemyevAV_2024}
Artemyev, A.~V., Zhang, X.~J., Demekhov, A.~G., {et~al.} 2024, Journal of Geophysical Research: Space Physics, 129, \dodoi{10.1029/2023ja032287}

\bibitem[{Asai {et~al.}(2002)Asai, Masuda, Yokoyama, Shimojo, Isobe, Kurokawa, \& Shibata}]{AsaiA_2002}
Asai, A., Masuda, S., Yokoyama, T., {et~al.} 2002, The Astrophysical Journal, 578, L91, \dodoi{10.1086/344566}

\bibitem[{Aschwanden(2005)}]{AschwandenM_2005}
Aschwanden, M.~J. 2005, Physics of the Solar Corona (Springer Berlin Heidelberg), \dodoi{10.1007/3-540-30766-4}

\bibitem[{Aschwanden(2012)}]{AschwandenM_2012}
---. 2012, Space Science Reviews, 171, 3, \dodoi{10.1007/s11214-011-9865-x}

\bibitem[{Aschwanden {et~al.}(1996)Aschwanden, Hudson, Kosugi, \& Schwartz}]{AschwandenMJ_1996}
Aschwanden, M.~J., Hudson, H., Kosugi, T., \& Schwartz, R.~A. 1996, Astrophys J, 464, 985, \dodoi{Doi 10.1086/177386}

\bibitem[{Atoyan \& Aharonian(1996)}]{AtoyanAM_1996}
Atoyan, A.~M., \& Aharonian, F.~A. 1996, Monthly Notices of the Royal Astronomical Society, 278, 525, \dodoi{10.1093/mnras/278.2.525}

\bibitem[{Bamba {et~al.}(2003)Bamba, Yamazaki, Ueno, \& Koyama}]{BambaA_2003}
Bamba, A., Yamazaki, R., Ueno, M., \& Koyama, K. 2003, The Astrophysical Journal, 589, 827, \dodoi{10.1086/374687}

\bibitem[{Bamba {et~al.}(2008)Bamba, Fukazawa, Hiraga, Hughes, Katagiri, Kokubun, Koyama, Miyata, Mizuno, Mori, Nakajima, Ozaki, Petre, Takahashi, Takahashi, Tanaka, Terada, Uchiyama, Watanabe, \& Yamaguchi}]{BambaA_2008}
Bamba, A., Fukazawa, Y., Hiraga, J.~S., {et~al.} 2008, Publications of the Astronomical Society of Japan, 60, S153, \dodoi{10.1093/pasj/60.sp1.S153}

\bibitem[{Band {et~al.}(1993)Band, Matteson, Ford, Schaefer, Palmer, Teegarden, Cline, Briggs, Paciesas, Pendleton, Fishman, Kouveliotou, Meegan, Wilson, \& Lestrade}]{BandD_1993}
Band, D., Matteson, J., Ford, L., {et~al.} 1993, The Astrophysical Journal, 413, \dodoi{10.1086/172995}

\bibitem[{Becker {et~al.}(2021)Becker, Alexander, Connerney, Brennan, Guillaume, Adumitroaie, Florence, Kollmann, Mauk, \& Bolton}]{BeckerHN_2021}
Becker, H.~N., Alexander, J.~W., Connerney, J. E.~P., {et~al.} 2021, Journal of Geophysical Research: Planets, 126, \dodoi{10.1029/2020je006772}

\bibitem[{Bell(2004)}]{BellAR_2004}
Bell, A.~R. 2004, Monthly Notices of the Royal Astronomical Society, 353, 550, \dodoi{10.1111/j.1365-2966.2004.08097.x}

\bibitem[{Berezhko {et~al.}(2002)Berezhko, Ksenofontov, \& Völk}]{BerezhkoEG_2002}
Berezhko, E.~G., Ksenofontov, L.~T., \& Völk, H.~J. 2002, Astronomy \& Astrophysics, 395, 943, \dodoi{10.1051/0004-6361:20021219}

\bibitem[{Bird \& Edenhofer(1990)}]{BirdMK_1990}
Bird, M.~K., \& Edenhofer, P. 1990, Remote Sensing Observations of the Solar Corona (Berlin Hidelberg: Springer-Verlag), 13--98

\bibitem[{Blake {et~al.}(1992)Blake, Kolasinski, Fillius, \& Mullen}]{BlakeJB_1992}
Blake, J.~B., Kolasinski, W.~A., Fillius, R.~W., \& Mullen, E.~G. 1992, Geophysical Research Letters, 19, 821, \dodoi{10.1029/92gl00624}

\bibitem[{Blandford \& Eichler(1987)}]{BlandfordR_1987}
Blandford, R., \& Eichler, D. 1987, Physics Reports, 154, 1, \dodoi{10.1016/0370-1573(87)90134-7}

\bibitem[{Bolton {et~al.}(2002)Bolton, Janssen, Thorne, Levin, Klein, Gulkis, Bastian, Sault, Elachi, Hofstadter, Bunker, Dulk, Gudim, Hamilton, Johnson, Leblanc, Liepack, McLeod, Roller, Roth, \& West}]{BoltonSJ_2002}
Bolton, S.~J., Janssen, M., Thorne, R., {et~al.} 2002, Nature, 415, 987, \dodoi{10.1038/415987a}

\bibitem[{Bowers {et~al.}(2024)Bowers, Jackman, Sun, Holmberg, Jia, \& Griton}]{BowersCF_2024}
Bowers, C.~F., Jackman, C.~M., Sun, W., {et~al.} 2024, Journal of Geophysical Research: Space Physics, 129, \dodoi{10.1029/2023ja032162}

\bibitem[{Buhler \& Blandford(2014)}]{BuhlerR_2014}
Buhler, R., \& Blandford, R. 2014, Rep Prog Phys, 77, 066901, \dodoi{10.1088/0034-4885/77/6/066901}

\bibitem[{Burlaga {et~al.}(2008)Burlaga, Ness, Acuña, Lepping, Connerney, \& Richardson}]{BurlagaLF_2008}
Burlaga, L.~F., Ness, N.~F., Acuña, M.~H., {et~al.} 2008, Nature, 454, 75, \dodoi{10.1038/nature07029}

\bibitem[{Carilli \& Barthel(1996)}]{CarilliCL_1996}
Carilli, C.~L., \& Barthel, P.~D. 1996, Astron Astrophys Rev, 7, 1, \dodoi{10.1007/s001590050001}

\bibitem[{Carmichael(1962)}]{CarmichaelH_1962}
Carmichael, H. 1962, Space Science Reviews, 1, \dodoi{10.1007/bf00174635}

\bibitem[{Casse \& Marcowith(2005)}]{CasseF_2005}
Casse, F., \& Marcowith, A. 2005, Astroparticle Physics, 23, 31, \dodoi{10.1016/j.astropartphys.2004.11.003}

\bibitem[{Chen {et~al.}(2015)Chen, Bastian, Shen, Gary, Krucker, \& Glesener}]{ChenB_2015}
Chen, Bastian, T.~S., Shen, C., {et~al.} 2015, Science, 350, 1238, \dodoi{10.1126/science.aac8467}

\bibitem[{Chen {et~al.}(2020)Chen, Shen, Gary, Reeves, Fleishman, Yu, Guo, Krucker, Lin, Nita, \& Kong}]{ChenB_2020}
Chen, B., Shen, C.~C., Gary, D.~E., {et~al.} 2020, Nature Astronomy, 4, 1140, \dodoi{10.1038/s41550-020-1147-7}

\bibitem[{Chien {et~al.}(2023)Chien, Gao, Zhang, Ji, Blackman, Daughton, Stanier, Le, Guo, Follett, Chen, Fiksel, Bleotu, Cauble, Chen, Fazzini, Flippo, French, Froula, Fuchs, Fujioka, Hill, Klein, Kuranz, Nilson, Rasmus, \& Takizawa}]{ChienA_2023}
Chien, A., Gao, L., Zhang, S., {et~al.} 2023, Nat Phys, \dodoi{10.1038/s41567-022-01839-x}

\bibitem[{Christon {et~al.}(1988)Christon, Mitchell, Williams, Frank, Huang, \& Eastman}]{ChristonSP_1988}
Christon, S.~P., Mitchell, D.~G., Williams, D.~J., {et~al.} 1988, Journal of Geophysical Research, 93, 2562, \dodoi{10.1029/JA093iA04p02562}

\bibitem[{Christon {et~al.}(1989)Christon, Williams, Mitchell, Frank, \& Huang}]{ChristonSP_1989}
Christon, S.~P., Williams, D.~J., Mitchell, D.~G., Frank, L.~A., \& Huang, C.~Y. 1989, Journal of Geophysical Research, 94, 13409, \dodoi{10.1029/JA094iA10p13409}

\bibitem[{Christon {et~al.}(1991)Christon, Williams, Mitchell, Huang, \& Frank}]{ChristonSP_1991}
Christon, S.~P., Williams, D.~J., Mitchell, D.~G., Huang, C.~Y., \& Frank, L.~A. 1991, Journal of Geophysical Research, 96, 1, \dodoi{10.1029/90ja01633}

\bibitem[{Cummings \& Stone(1998)}]{CummingsAC_1998}
Cummings, A.~C., \& Stone, E.~C. 1998, Space Science Reviews, 83, 51, \dodoi{10.1023/a:1005057010311}

\bibitem[{de~Pater \& Dunn(2003)}]{dePaterI_2003}
de~Pater, I., \& Dunn, D.~E. 2003, Icarus, 163, 449, \dodoi{10.1016/s0019-1035(03)00068-x}

\bibitem[{Dewey {et~al.}(2018)Dewey, Raines, Sun, Slavin, \& Poh}]{DeweyRM_2018}
Dewey, R.~M., Raines, J.~M., Sun, W., Slavin, J.~A., \& Poh, G. 2018, Geophysical Research Letters, 45, \dodoi{10.1029/2018gl079056}

\bibitem[{Diesing(2023)}]{DiesingR_2023}
Diesing, R. 2023, The Astrophysical Journal, 958, \dodoi{10.3847/1538-4357/ad00b1}

\bibitem[{Dmitruk {et~al.}(2003)Dmitruk, Matthaeus, Seenu, \& Brown}]{DmitrukP_2003}
Dmitruk, P., Matthaeus, W.~H., Seenu, N., \& Brown, M.~R. 2003, The Astrophysical Journal, 597, L81, \dodoi{10.1086/379751}

\bibitem[{Drake {et~al.}(2010)Drake, Opher, Swisdak, \& Chamoun}]{DrakeJF_2010}
Drake, J.~F., Opher, M., Swisdak, M., \& Chamoun, J.~N. 2010, The Astrophysical Journal, 709, 963, \dodoi{10.1088/0004-637x/709/2/963}

\bibitem[{Drury(2012)}]{DruryLOC_2012}
Drury, L.~O. 2012, Astroparticle Physics, 39-40, 52, \dodoi{10.1016/j.astropartphys.2012.02.006}

\bibitem[{Ellison \& Ramaty(1985)}]{EllisonDC_1985}
Ellison, D.~C., \& Ramaty, R. 1985, The Astrophysical Journal, 298, \dodoi{10.1086/163623}

\bibitem[{Evenson {et~al.}(1984)Evenson, Meyer, Yanagita, \& Forrest}]{EvensonP_1984}
Evenson, P., Meyer, P., Yanagita, S., \& Forrest, D.~J. 1984, The Astrophysical Journal, 283, \dodoi{10.1086/162323}

\bibitem[{Fairfield(1971)}]{FairfieldDH_1971}
Fairfield, D.~H. 1971, Journal of Geophysical Research, 76, 6700, \dodoi{10.1029/JA076i028p06700}

\bibitem[{Fairfield \& Jones(1996)}]{FairfieldDH_1996}
Fairfield, D.~H., \& Jones, J. 1996, Journal of Geophysical Research: Space Physics, 101, 7785, \dodoi{10.1029/95ja03713}

\bibitem[{Feigelson {et~al.}(1995)Feigelson, Laurent-Muehleisen, Kollgaard, \& Fomalont}]{FeigelsonED_1995}
Feigelson, E.~D., Laurent-Muehleisen, S.~A., Kollgaard, R.~I., \& Fomalont, E.~B. 1995, The Astrophysical Journal, 449, \dodoi{10.1086/309642}

\bibitem[{Fischer {et~al.}(1996)Fischer, Pehlke, Wibberenz, Lanzerotti, \& Mihalov}]{FischerHM_1996}
Fischer, H.~M., Pehlke, E., Wibberenz, G., Lanzerotti, L.~J., \& Mihalov, J.~D. 1996, Science, 272, 856, \dodoi{10.1126/science.272.5263.856}

\bibitem[{Fiuza {et~al.}(2020)Fiuza, Swadling, Grassi, Rinderknecht, Higginson, Ryutov, Bruulsema, Drake, Funk, Glenzer, Gregori, Li, Pollock, Remington, Ross, Rozmus, Sakawa, Spitkovsky, Wilks, \& Park}]{FiuzaF_2020}
Fiuza, F., Swadling, G.~F., Grassi, A., {et~al.} 2020, Nat Phys, 16, 916, \dodoi{10.1038/s41567-020-0919-4}

\bibitem[{Fukui {et~al.}(2021)Fukui, Sano, Yamane, Hayakawa, Inoue, Tachihara, Rowell, \& Einecke}]{FukuiY_2021}
Fukui, Y., Sano, H., Yamane, Y., {et~al.} 2021, The Astrophysical Journal, 915, \dodoi{10.3847/1538-4357/abff4a}

\bibitem[{Gardner \& Milne(1965)}]{GardnerFF_1965}
Gardner, F.~F., \& Milne, D.~K. 1965, The Astronomical Journal, 70, \dodoi{10.1086/109813}

\bibitem[{Gershman {et~al.}(2024)Gershman, Fuselier, Cohen, Turner, Liu, Chen, Phan, Stawarz, Dibraccio, Masters, Ebert, Sun, Harada, \& Swisdak}]{GershmanD_2024}
Gershman, D.~J., Fuselier, S.~A., Cohen, I.~J., {et~al.} 2024, Space Science Reviews, 220, \dodoi{10.1007/s11214-023-01017-2}

\bibitem[{Giacalone {et~al.}(2022)Giacalone, Fahr, Fichtner, Florinski, Heber, Hill, Kóta, Leske, Potgieter, \& Rankin}]{GiacaloneJ_2022}
Giacalone, J., Fahr, H., Fichtner, H., {et~al.} 2022, Space Science Reviews, 218, \dodoi{10.1007/s11214-022-00890-7}

\bibitem[{Goldreich \& Julian(1969)}]{GoldreichP_1969}
Goldreich, P., \& Julian, W.~H. 1969, The Astrophysical Journal, 157, \dodoi{10.1086/150119}

\bibitem[{Goldstein {et~al.}(1986)Goldstein, Matthaeus, \& Ambrosiano}]{GoldsteinML_1986}
Goldstein, M.~L., Matthaeus, W.~H., \& Ambrosiano, J.~J. 1986, Geophysical Research Letters, 13, 205, \dodoi{10.1029/GL013i003p00205}

\bibitem[{Green(2019)}]{GreenDA_2019}
Green, D.~A. 2019, Journal of Astrophysics and Astronomy, 40, \dodoi{10.1007/s12036-019-9601-6}

\bibitem[{Guo {et~al.}(2024)Guo, Liu, Zenitani, \& Hoshino}]{GuoFan_2024}
Guo, F., Liu, Y.-H., Zenitani, S., \& Hoshino, M. 2024, Space Science Reviews, 220, \dodoi{10.1007/s11214-024-01073-2}

\bibitem[{Guo {et~al.}(2023)Guo, Li, Zhang, Dobynde, Wang, Xu, Berger, Semkova, Wimmer‐Schweingruber, Hassler, Zeitlin, Ehresmann, Matthiä, \& Zhuang}]{GuoJingnan_2023}
Guo, J., Li, X., Zhang, J., {et~al.} 2023, Geophysical Research Letters, 50, \dodoi{10.1029/2023gl103069}

\bibitem[{Guo \& Yao(2024)}]{GuoRuilong_2024}
Guo, R., \& Yao, Z. 2024, Reviews of Modern Plasma Physics, 8, \dodoi{10.1007/s41614-024-00162-7}

\bibitem[{Hairston {et~al.}(2005)Hairston, Drake, \& Skoug}]{HairstonMR_2005}
Hairston, M.~R., Drake, K.~A., \& Skoug, R. 2005, Journal of Geophysical Research: Space Physics, 110, \dodoi{10.1029/2004ja010864}

\bibitem[{Hess(1959)}]{HessWN_1959}
Hess, W.~N. 1959, Physical Review Letters, 3, 11, \dodoi{10.1103/PhysRevLett.3.11}

\bibitem[{Hill {et~al.}(2008)Hill, Thomsen, Henderson, Tokar, Coates, McAndrews, Lewis, Mitchell, Jackman, Russell, Dougherty, Crary, \& Young}]{HillTW_2008}
Hill, T.~W., Thomsen, M.~F., Henderson, M.~G., {et~al.} 2008, Journal of Geophysical Research: Space Physics, 113, \dodoi{10.1029/2007ja012626}

\bibitem[{Hillas(1984)}]{HillasAM_1984}
Hillas, a.~M. 1984, Annual Review of Astronomy and Astrophysics, 22, 425, \dodoi{10.1146/annurev.aa.22.090184.002233}

\bibitem[{Hillas(2005)}]{HillasAM_2005}
Hillas, A.~M. 2005, Journal of Physics G: Nuclear and Particle Physics, 31, R95, \dodoi{10.1088/0954-3899/31/5/r02}

\bibitem[{Ho {et~al.}(2011)Ho, Krimigis, Gold, Baker, Slavin, Anderson, Korth, Starr, Lawrence, McNutt, \& Solomon}]{HoG_2011_Science}
Ho, G.~C., Krimigis, S.~M., Gold, R.~E., {et~al.} 2011, Science, 333, 1865, \dodoi{10.1126/science.1211141}

\bibitem[{Ho {et~al.}(2012)Ho, Krimigis, Gold, Baker, Anderson, Korth, Slavin, McNutt, Winslow, \& Solomon}]{HoG_2012}
---. 2012, Journal of Geophysical Research: Space Physics, 117, \dodoi{10.1029/2012ja017983}

\bibitem[{Jackman {et~al.}(2019)Jackman, Thomsen, \& Dougherty}]{JackmanCM_2019}
Jackman, C.~M., Thomsen, M.~F., \& Dougherty, M.~K. 2019, Journal of Geophysical Research: Space Physics, 124, 8865, \dodoi{10.1029/2019ja026628}

\bibitem[{Jackman {et~al.}(2014)Jackman, Arridge, André, Bagenal, Birn, Freeman, Jia, Kidder, Milan, Radioti, Slavin, Vogt, Volwerk, \& Walsh}]{JackmanCM_2014}
Jackman, C.~M., Arridge, C.~S., André, N., {et~al.} 2014, Space Science Reviews, 182, 85, \dodoi{10.1007/s11214-014-0060-8}

\bibitem[{Jaynes {et~al.}(2015)Jaynes, Baker, Singer, Rodriguez, Loto'aniu, Ali, Elkington, Li, Kanekal, Claudepierre, Fennell, Li, Thorne, Kletzing, Spence, \& Reeves}]{JaynesAN_2015}
Jaynes, A.~N., Baker, D.~N., Singer, H.~J., {et~al.} 2015, Journal of Geophysical Research: Space Physics, 120, 7240, \dodoi{10.1002/2015ja021234}

\bibitem[{Jokipii(1987)}]{JokipiiJR_1987}
Jokipii, J.~R. 1987, Astrophys J, 313, 842, \dodoi{Doi 10.1086/165022}

\bibitem[{Jokipii \& Giacalone(1998)}]{JokipiiJR_1998}
Jokipii, J.~R., \& Giacalone, J. 1998, Space Science Reviews, 83, 123, \dodoi{10.1023/a:1005077629875}

\bibitem[{Kahler(1994)}]{KahlerS_1994}
Kahler, S. 1994, The Astrophysical Journal, 428, \dodoi{10.1086/174292}

\bibitem[{Kaneda {et~al.}(1995)Kaneda, Tashiro, Ikebe, Ishisaki, Kubo, Makishima, Ohashi, Saito, Tabara, \& Takahashi}]{KanedaH_1995}
Kaneda, H., Tashiro, M., Ikebe, Y., {et~al.} 1995, The Astrophysical Journal, 453, \dodoi{10.1086/309742}

\bibitem[{Katsuda {et~al.}(2009)Katsuda, Petre, Long, Reynolds, Winkler, Mori, \& Tsunemi}]{KatsudaS_2009}
Katsuda, S., Petre, R., Long, K.~S., {et~al.} 2009, The Astrophysical Journal, 692, L105, \dodoi{10.1088/0004-637x/692/2/l105}

\bibitem[{Kennel \& Coroniti(1984)}]{KennelCF_1984_Confinement}
Kennel, C.~F., \& Coroniti, F.~V. 1984, The Astrophysical Journal, 283, \dodoi{10.1086/162356}

\bibitem[{Kivelson \& Khurana(2002)}]{KivelsonM_2002}
Kivelson, M.~G., \& Khurana, K.~K. 2002, Journal of Geophysical Research: Space Physics, 107, \dodoi{10.1029/2001ja000249}

\bibitem[{Kliem(1994)}]{KliemB_1994}
Kliem, B. 1994, Astrophys J Suppl S, 90, 719, \dodoi{10.1086/191896}

\bibitem[{Kollmann {et~al.}(2017)Kollmann, Paranicas, Clark, Mauk, Haggerty, Rymer, Santos‐Costa, Connerney, Allegrini, Valek, Kurth, Gladstone, Levin, \& Bolton}]{KollmannP_2017}
Kollmann, P., Paranicas, C., Clark, G., {et~al.} 2017, Geophysical Research Letters, 44, 5259, \dodoi{10.1002/2017gl073730}

\bibitem[{Kotera \& Olinto(2011)}]{KoteraK_2011}
Kotera, K., \& Olinto, A.~V. 2011, Annual Review of Astronomy and Astrophysics, 49, 119, \dodoi{10.1146/annurev-astro-081710-102620}

\bibitem[{Koyama {et~al.}(1995)Koyama, Petre, Gotthelf, Hwang, Matsuura, Ozaki, \& Holt}]{KoyamaK_1995}
Koyama, K., Petre, R., Gotthelf, E.~V., {et~al.} 1995, Nature, 378, 255, \dodoi{DOI 10.1038/378255a0}

\bibitem[{Krimigis(1979)}]{KrimigisSM_1979}
Krimigis, S.~M. 1979, AIP Conference Proceedings, 56, 179, \dodoi{10.1063/1.32079}

\bibitem[{Kronberg {et~al.}(2008)Kronberg, Woch, Krupp, \& Lagg}]{KronbergEA_2008}
Kronberg, E.~A., Woch, J., Krupp, N., \& Lagg, A. 2008, Journal of Geophysical Research: Space Physics, 113, \dodoi{10.1029/2008ja013332}

\bibitem[{Krucker {et~al.}(2008)Krucker, Battaglia, Cargill, Fletcher, Hudson, MacKinnon, Masuda, Sui, Tomczak, Veronig, Vlahos, \& White}]{KruckerS_2008_review}
Krucker, S., Battaglia, M., Cargill, P.~J., {et~al.} 2008, Astron Astrophys Rev, 16, 155, \dodoi{10.1007/s00159-008-0014-9}

\bibitem[{Krupp {et~al.}(2018)Krupp, Roussos, Kollmann, Mitchell, Paranicas, Krimigis, Hamilton, Hedman, \& Dougherty}]{KruppN_2018}
Krupp, N., Roussos, E., Kollmann, P., {et~al.} 2018, Geophysical Research Letters, 45, \dodoi{10.1029/2018gl078096}

\bibitem[{Lagage \& Cesarsky(1983)}]{LagagePO_1983}
Lagage, P.~O., \& Cesarsky, C.~J. 1983, Astronomy \& Astrophysics, 125, 249

\bibitem[{Larrodera \& Cid(2020)}]{LarroderaC_2020}
Larrodera, C., \& Cid, C. 2020, Astronomy \& Astrophysics, 635, \dodoi{10.1051/0004-6361/201937307}

\bibitem[{Lawrence {et~al.}(2015)Lawrence, Anderson, Baker, Feldman, Ho, Korth, McNutt, Peplowski, Solomon, Starr, Vandegriff, \& Winslow}]{LawrenceDJ_2015}
Lawrence, D.~J., Anderson, B.~J., Baker, D.~N., {et~al.} 2015, Journal of Geophysical Research: Space Physics, 120, 2851, \dodoi{10.1002/2014ja020792}

\bibitem[{Li {et~al.}(2018)Li, Ballet, Miceli, Zhou, Vink, Chen, Acero, Decourchelle, \& Bregman}]{LiJT_2018}
Li, J.-T., Ballet, J., Miceli, M., {et~al.} 2018, The Astrophysical Journal, 864, \dodoi{10.3847/1538-4357/aad598}

\bibitem[{Li {et~al.}(2023)Li, Yue, Liu, Zong, Zou, \& Ye}]{LiYuXuan_2023}
Li, Y., Yue, C., Liu, Y., {et~al.} 2023, Earth and Planetary Physics, 7, 109, \dodoi{10.26464/epp2023009}

\bibitem[{Lin {et~al.}(2002)Lin, Dennis, Hurford, Smith, Zehnder, Harvey, Curtis, Pankow, Turin, Bester, Csillaghy, Lewis, Madden, van Beek, Appleby, Raudorf, McTiernan, Ramaty, Schmahl, Schwartz, Krucker, Abiad, Quinn, Berg, Hashii, Sterling, Jackson, Pratt, Campbell, Malone, Landis, Barrington-Leigh, Slassi-Sennou, Cork, Clark, Amato, Orwig, Boyle, Banks, Shirey, Tolbert, Zarro, Snow, Thomsen, Henneck, McHedlishvili, Ming, Fivian, Jordan, Wanner, Crubb, Preble, Matranga, Benz, Hudson, Canfield, Holman, Crannell, Kosugi, Emslie, Vilmer, Brown, Johns-Krull, Aschwanden, Metcalf, \& Conway}]{LinRP_2002}
Lin, R.~P., Dennis, B.~R., Hurford, G.~J., {et~al.} 2002, Solar Physics, 210, 3, \dodoi{10.1023/a:1022428818870}

\bibitem[{Liu {et~al.}(2020)Liu, Wang, Wimmer‐Schweingruber, Krucker, \& Mason}]{LiuZixuan_2020}
Liu, Z., Wang, L., Wimmer‐Schweingruber, R.~F., Krucker, S., \& Mason, G.~M. 2020, Journal of Geophysical Research: Space Physics, 125, \dodoi{10.1029/2020ja028702}

\bibitem[{Longair(2011)}]{LongairM_2011}
Longair, M.~S. 2011, High Energy Astrophysics, 3rd edn. (United Kingdom: Cambridge University Press), 861, \dodoi{10.1017/cbo9780511778346}

\bibitem[{Lovell {et~al.}(1998)Lovell, Duldig, \& Humble}]{LovellJL_1998}
Lovell, J.~L., Duldig, M.~L., \& Humble, J.~E. 1998, Journal of Geophysical Research: Space Physics, 103, 23733, \dodoi{10.1029/98ja02100}

\bibitem[{Lu {et~al.}(2021)Lu, Guo, Kilian, Li, Huang, \& Liang}]{LuYingchao_2021}
Lu, Y., Guo, F., Kilian, P., {et~al.} 2021, The Astrophysical Journal, 908, \dodoi{10.3847/1538-4357/abd406}

\bibitem[{Makishima(1999)}]{MakishimaK_1999}
Makishima, K. 1999, Astron Nachr, 320, 163.
\newblock \url{<Go to ISI>://WOS:000084277700001}

\bibitem[{Marsden {et~al.}(1984)Marsden, Gillett, Jennings, Emerson, de~Jong, \& Olnon}]{MarsdenPL_1984}
Marsden, P.~L., Gillett, F.~C., Jennings, R.~E., {et~al.} 1984, The Astrophysical Journal, 278, \dodoi{10.1086/184215}

\bibitem[{Martens(1988)}]{MartensPCH_1988}
Martens, P. C.~H. 1988, The Astrophysical Journal, 330, \dodoi{10.1086/185220}

\bibitem[{Masters {et~al.}(2013)Masters, Stawarz, Fujimoto, Schwartz, Sergis, Thomsen, Retinò, Hasegawa, Zieger, Lewis, Coates, Canu, \& Dougherty}]{MastersA_2013}
Masters, A., Stawarz, L., Fujimoto, M., {et~al.} 2013, Nat Phys, 9, 164, \dodoi{10.1038/nphys2541}

\bibitem[{Masuda {et~al.}(1994)Masuda, Kosugi, Hara, Tsuneta, \& Ogawara}]{MasudaS_1994}
Masuda, S., Kosugi, T., Hara, H., Tsuneta, S., \& Ogawara, Y. 1994, Nature, 371, 495, \dodoi{10.1038/371495a0}

\bibitem[{Matthaeus {et~al.}(1984)Matthaeus, Ambrosiano, \& Goldstein}]{MatthaeusWH_1984}
Matthaeus, W.~H., Ambrosiano, J.~J., \& Goldstein, M.~L. 1984, Physical Review Letters, 53, 1449, \dodoi{10.1103/PhysRevLett.53.1449}

\bibitem[{Matthews {et~al.}(2020)Matthews, Bell, \& Blundell}]{MatthewsJH_2020}
Matthews, J.~H., Bell, A.~R., \& Blundell, K.~M. 2020, New Astronomy Reviews, 89, \dodoi{10.1016/j.newar.2020.101543}

\bibitem[{Mazur {et~al.}(2023)Mazur, O’Brien, \& Looper}]{MazurJE_2023}
Mazur, J.~E., O’Brien, T.~P., \& Looper, M.~D. 2023, Space Science Reviews, 219, \dodoi{10.1007/s11214-023-00962-2}

\bibitem[{McAndrews {et~al.}(2009)McAndrews, Thomsen, Arridge, Jackman, Wilson, Henderson, Tokar, Khurana, Sittler, Coates, \& Dougherty}]{McAndrewsHJ_2009}
McAndrews, H.~J., Thomsen, M.~F., Arridge, C.~S., {et~al.} 2009, Planetary and Space Science, 57, 1714, \dodoi{10.1016/j.pss.2009.03.003}

\bibitem[{McCracken {et~al.}(2023)McCracken, Shea, \& Smart}]{McCrackenKG_2023}
McCracken, K.~G., Shea, M.~A., \& Smart, D.~F. 2023, Advances in Space Research, 72, 3414, \dodoi{10.1016/j.asr.2023.06.049}

\bibitem[{Meisenheimer {et~al.}(1997)Meisenheimer, Yates, \& Röser}]{MeisenheimerK_1997}
Meisenheimer, K., Yates, M.~G., \& Röser, H.~J. 1997, AstroAstronomy \& Astrophysics, 325, 57

\bibitem[{Mewaldt {et~al.}(2012)Mewaldt, Looper, Cohen, Haggerty, Labrador, Leske, Mason, Mazur, \& von Rosenvinge}]{MewaldtRA_2012a}
Mewaldt, R.~A., Looper, M.~D., Cohen, C. M.~S., {et~al.} 2012, Space Science Reviews, 171, 97, \dodoi{10.1007/s11214-012-9884-2}

\bibitem[{Miyashita {et~al.}(2009)Miyashita, Machida, Kamide, Nagata, Liou, Fujimoto, Ieda, Saito, Russell, Christon, Nosé, Frey, Shinohara, Mukai, Saito, \& Hayakawa}]{MiyashitaY_2009}
Miyashita, Y., Machida, S., Kamide, Y., {et~al.} 2009, J Geophys Res-Space, 114, n/a, \dodoi{Artn A01211 10.1029/2008ja013225}

\bibitem[{Mori {et~al.}(2022)Mori, Hailey, Bridges, Mandel, Garvin, Grefenstette, Dunn, Hord, Branduardi-Raymont, Clarke, Jackman, Nynka, \& Ray}]{MoriK_2022}
Mori, K., Hailey, C., Bridges, G., {et~al.} 2022, Nature Astronomy, 6, 442, \dodoi{10.1038/s41550-021-01594-8}

\bibitem[{Moses {et~al.}(1989)Moses, Droege, Meyer, \& Evenson}]{MosesD_1989}
Moses, D., Droege, W., Meyer, P., \& Evenson, P. 1989, The Astrophysical Journal, 346, \dodoi{10.1086/168034}

\bibitem[{Norman {et~al.}(1995)Norman, Melrose, \& Achterberg}]{NormanCA_1995}
Norman, C.~A., Melrose, D.~B., \& Achterberg, A. 1995, The Astrophysical Journal, 454, \dodoi{10.1086/176465}

\bibitem[{Oka {et~al.}(2015)Oka, Krucker, Hudson, \& Saint-Hilaire}]{OkaM_2015}
Oka, M., Krucker, S., Hudson, H.~S., \& Saint-Hilaire, P. 2015, The Astrophysical Journal, 799, 129, \dodoi{10.1088/0004-637x/799/2/129}

\bibitem[{Oka {et~al.}(2010)Oka, Phan, Krucker, Fujimoto, \& Shinohara}]{OkaM_2010_coa}
Oka, M., Phan, T.~D., Krucker, S., Fujimoto, M., \& Shinohara, I. 2010, The Astrophysical Journal, 714, 915, \dodoi{10.1088/0004-637x/714/1/915}

\bibitem[{Oka {et~al.}(2006)Oka, Terasawa, Seki, Fujimoto, Kasaba, Kojima, Shinohara, Matsui, Matsumoto, Saito, \& Mukai}]{OkaM_2006}
Oka, M., Terasawa, T., Seki, Y., {et~al.} 2006, Geophysical Research Letters, 33, L24104, \dodoi{10.1029/2006gl028156}

\bibitem[{Oka {et~al.}(2018)Oka, Birn, Battaglia, Chaston, Hatch, Livadiotis, Imada, Miyoshi, Kuhar, Effenberger, Eriksson, Khotyaintsev, \& Retinò}]{OkaM_2018}
Oka, M., Birn, J., Battaglia, M., {et~al.} 2018, Space Science Reviews, 214, 82, \dodoi{10.1007/s11214-018-0515-4}

\bibitem[{Oka {et~al.}(2023)Oka, Birn, Egedal, Guo, Ergun, Turner, Khotyaintsev, Hwang, Cohen, \& Drake}]{OkaM_2023}
Oka, M., Birn, J., Egedal, J., {et~al.} 2023, Space Science Reviews, 219, \dodoi{10.1007/s11214-023-01011-8}

\bibitem[{Paranicas {et~al.}(2010)Paranicas, Mitchell, Krimigis, Carbary, Brandt, Turner, Roussos, Krupp, Kivelson, Khurana, Cooper, Armstrong, \& Burton}]{ParanicasC_2010}
Paranicas, C., Mitchell, D.~G., Krimigis, S.~M., {et~al.} 2010, Journal of Geophysical Research: Space Physics, 115, \dodoi{10.1029/2009ja014971}

\bibitem[{Parizot {et~al.}(2006)Parizot, Marcowith, Ballet, \& Gallant}]{ParizotE_2006}
Parizot, E., Marcowith, A., Ballet, J., \& Gallant, Y.~A. 2006, Astronomy \& Astrophysics, 453, 387, \dodoi{10.1051/0004-6361:20064985}

\bibitem[{Phan {et~al.}(2018)Phan, Eastwood, Shay, Drake, Sonnerup, Fujimoto, Cassak, Øieroset, Burch, Torbert, Rager, Dorelli, Gershman, Pollock, Pyakurel, Haggerty, Khotyaintsev, Lavraud, Saito, Oka, Ergun, Retino, Le~Contel, Argall, Giles, Moore, Wilder, Strangeway, Russell, Lindqvist, \& Magnes}]{PhanTD_2018}
Phan, T.~D., Eastwood, J.~P., Shay, M.~A., {et~al.} 2018, Nature, 557, 202, \dodoi{10.1038/s41586-018-0091-5}

\bibitem[{Poh {et~al.}(2017)Poh, Slavin, Jia, Raines, Imber, Sun, Gershman, DiBraccio, Genestreti, \& Smith}]{PohG_2017}
Poh, G., Slavin, J.~A., Jia, X., {et~al.} 2017, Geophysical Research Letters, 44, 678, \dodoi{10.1002/2016gl071612}

\bibitem[{Ptitsyna \& Troitsky(2010)}]{PtitsynaKV_2010}
Ptitsyna, K.~V., \& Troitsky, S.~V. 2010, Physics-Uspekhi, 53, 691, \dodoi{10.3367/UFNe.0180.201007c.0723}

\bibitem[{Rachen \& Biermann(1993)}]{RachenJP_1993}
Rachen, J.~P., \& Biermann, P.~L. 1993, Astronomy \& Astrophysics, 272, 161

\bibitem[{Raines {et~al.}(2011)Raines, Slavin, Zurbuchen, Gloeckler, Anderson, Baker, Korth, Krimigis, \& McNutt}]{RainesJM_2011}
Raines, J.~M., Slavin, J.~A., Zurbuchen, T.~H., {et~al.} 2011, Planetary and Space Science, 59, 2004, \dodoi{10.1016/j.pss.2011.02.004}

\bibitem[{Roussos {et~al.}(2014)Roussos, Krupp, Paranicas, Carbary, Kollmann, Krimigis, \& Mitchell}]{RoussosE_2014}
Roussos, E., Krupp, N., Paranicas, C., {et~al.} 2014, Planetary and Space Science, 104, 3, \dodoi{10.1016/j.pss.2014.03.021}

\bibitem[{Roussos {et~al.}(2018)Roussos, Kollmann, Krupp, Kotova, Regoli, Paranicas, Mitchell, Krimigis, Hamilton, Brandt, Carbary, Christon, Dialynas, Dandouras, Hill, Ip, Jones, Livi, Mauk, Palmaerts, Roelof, Rymer, Sergis, \& Smith}]{RoussosE_2018}
Roussos, E., Kollmann, P., Krupp, N., {et~al.} 2018, Science, 362, 47, \dodoi{ARTN eaat1962 10.1126/science.aat1962}

\bibitem[{Saint-Hilaire \& Benz(2005)}]{SaintHilaireP_2005}
Saint-Hilaire, P., \& Benz, A.~O. 2005, Astronomy \& Astrophysics, 435, 743, \dodoi{10.1051/0004-6361:20041918}

\bibitem[{Scholer {et~al.}(1981)Scholer, Hovestadt, Ipavich, \& Gloeckler}]{ScholerM_1981}
Scholer, M., Hovestadt, D., Ipavich, F.~M., \& Gloeckler, G. 1981, Journal of Geophysical Research: Space Physics, 86, 9040, \dodoi{10.1029/JA086iA11p09040}

\bibitem[{Shepherd(2007)}]{ShepherdSG_2007}
Shepherd, S.~G. 2007, Journal of Atmospheric and Solar-Terrestrial Physics, 69, 234, \dodoi{10.1016/j.jastp.2006.07.022}

\bibitem[{Shi {et~al.}(2024)Shi, Feng, Chen, Ying, Li, Li, Li, Li, Ji, Huang, Li, Li, Zhao, Lu, Xue, Zhang, Song, Tian, Su, Zhang, Ge, Shan, Zhou, Tian, Li, Liu, Jing, Lei, \& Gan}]{ShiG_2024}
Shi, G., Feng, L., Chen, J., {et~al.} 2024, Solar Physics, 299, \dodoi{10.1007/s11207-024-02349-0}

\bibitem[{Shibata {et~al.}(1995)Shibata, Masuda, Shimojo, Hara, Yokoyama, Tsuneta, Kosugi, \& Ogawara}]{ShibataK_1995}
Shibata, K., Masuda, S., Shimojo, M., {et~al.} 1995, The Astrophysical Journal, 451, \dodoi{10.1086/309688}

\bibitem[{Simpson {et~al.}(1974)Simpson, Eraker, Lamport, \& Walpole}]{SimpsonJA_1974}
Simpson, J.~A., Eraker, J.~H., Lamport, J.~E., \& Walpole, P.~H. 1974, Science, 185, 160, \dodoi{10.1126/science.185.4146.160}

\bibitem[{Singer(1958{\natexlab{a}})}]{SingerSF_1958a}
Singer, S.~F. 1958{\natexlab{a}}, Physical Review Letters, 1, 171, \dodoi{10.1103/PhysRevLett.1.171}

\bibitem[{Singer(1958{\natexlab{b}})}]{SingerSF_1958b}
---. 1958{\natexlab{b}}, Physical Review Letters, 1, 181, \dodoi{10.1103/PhysRevLett.1.181}

\bibitem[{Slavin {et~al.}(2012)Slavin, Imber, Boardsen, DiBraccio, Sundberg, Sarantos, Nieves‐Chinchilla, Szabo, Anderson, Korth, Zurbuchen, Raines, Johnson, Winslow, Killen, McNutt, \& Solomon}]{SlavinJA_2012}
Slavin, J.~A., Imber, S.~M., Boardsen, S.~A., {et~al.} 2012, Journal of Geophysical Research: Space Physics, 117, \dodoi{10.1029/2012ja017926}

\bibitem[{Smart \& Shea(1985)}]{SmartDF_1985}
Smart, D.~F., \& Shea, M.~A. 1985, J Geophys Res-Space, 90, 183, \dodoi{DOI 10.1029/JA090iA01p00183}

\bibitem[{Smith {et~al.}(2016)Smith, Jackman, \& Thomsen}]{SmithAW_2016}
Smith, A.~W., Jackman, C.~M., \& Thomsen, M.~F. 2016, J Geophys Res Space Phys, 121, 2984, \dodoi{10.1002/2015JA022005}

\bibitem[{Sorathia {et~al.}(2018)Sorathia, Ukhorskiy, Merkin, Fennell, \& Claudepierre}]{SorathiaK_2018}
Sorathia, K.~A., Ukhorskiy, A.~Y., Merkin, V.~G., Fennell, J.~F., \& Claudepierre, S.~G. 2018, Journal of Geophysical Research: Space Physics, 123, 5590, \dodoi{10.1029/2018ja025506}

\bibitem[{Steenbrugge \& Blundell(2008)}]{SteenbruggeKC_2008}
Steenbrugge, K.~C., \& Blundell, K.~M. 2008, Monthly Notices of the Royal Astronomical Society, 388, 1457, \dodoi{10.1111/j.1365-2966.2007.12665.x}

\bibitem[{Suzuki {et~al.}(2022)Suzuki, Bamba, Yamazaki, \& Ohira}]{SuzukiH_2022}
Suzuki, H., Bamba, A., Yamazaki, R., \& Ohira, Y. 2022, The Astrophysical Journal, 924, \dodoi{10.3847/1538-4357/ac33b5}

\bibitem[{Takahashi {et~al.}(2008)Takahashi, Tanaka, Uchiyama, Hiraga, Nakazawa, Watanabe, Bamba, Hughes, Katagiri, Kataoka, Kokubun, Koyama, Mori, Petre, Takahashi, \& Tsuboi}]{TakahashiT_2008}
Takahashi, T., Tanaka, T., Uchiyama, Y., {et~al.} 2008, Publications of the Astronomical Society of Japan, 60, S131, \dodoi{10.1093/pasj/60.sp1.S131}

\bibitem[{Tanaka {et~al.}(2020)Tanaka, Uchida, Sano, \& Tsuru}]{TanakaT_2020}
Tanaka, T., Uchida, H., Sano, H., \& Tsuru, T.~G. 2020, The Astrophysical Journal Letters, 900, \dodoi{10.3847/2041-8213/abaef0}

\bibitem[{Tao {et~al.}(2024)Tao, Kataoka, \& Tanaka}]{TaoM_2024}
Tao, M., Kataoka, J., \& Tanaka, T. 2024, The Astrophysical Journal Letters, 970, \dodoi{10.3847/2041-8213/ad60c7}

\bibitem[{Terasawa(2001)}]{TerasawaT_2001}
Terasawa, T. 2001, Science and Technology of Advanced Materials, 2, 461, \dodoi{10.1016/s1468-6996(01)00144-9}

\bibitem[{Terasawa(2011)}]{TerasawaT_2011}
---. 2011, Proceedings of the International Astronomical Union, 6, 214, \dodoi{10.1017/s174392131100696x}

\bibitem[{Terasawa \& Nishida(1976)}]{TerasawaT_1976}
Terasawa, T., \& Nishida, A. 1976, Planetary and Space Science, 24, 855, \dodoi{10.1016/0032-0633(76)90076-3}

\bibitem[{Thomsen {et~al.}(2019)Thomsen, Jackman, \& Lamy}]{ThomsenMF_2019}
Thomsen, M.~F., Jackman, C.~M., \& Lamy, L. 2019, Journal of Geophysical Research: Space Physics, 124, 7799, \dodoi{10.1029/2019ja026819}

\bibitem[{Trattner {et~al.}(2023)Trattner, Fuselier, Schwartz, Kucharek, Burch, Ergun, Petrinec, \& Madanian}]{TrattnerKJ_2023}
Trattner, K.~J., Fuselier, S.~A., Schwartz, S.~J., {et~al.} 2023, Journal of Geophysical Research: Space Physics, 128, \dodoi{10.1029/2022ja030631}

\bibitem[{Tsuji \& Uchiyama(2016)}]{TsujiN_2016}
Tsuji, N., \& Uchiyama, Y. 2016, Publications of the Astronomical Society of Japan, \dodoi{10.1093/pasj/psw102}

\bibitem[{Tsuji {et~al.}(2019)Tsuji, Uchiyama, Aharonian, Berge, Higurashi, Krivonos, \& Tanaka}]{TsujiN_2019}
Tsuji, N., Uchiyama, Y., Aharonian, F., {et~al.} 2019, The Astrophysical Journal, 877, \dodoi{10.3847/1538-4357/ab1b29}

\bibitem[{Tsuneta \& Naito(1998)}]{TsunetaNaito_1998}
Tsuneta, S., \& Naito, T. 1998, The Astrophysical Journal, 495, L67, \dodoi{10.1086/311207}

\bibitem[{Turner {et~al.}(2018)Turner, Wilson, Liu, Cohen, Schwartz, Osmane, Fennell, Clemmons, Blake, Westlake, Mauk, Jaynes, Leonard, Baker, Strangeway, Russell, Gershman, Avanov, Giles, Torbert, Broll, Gomez, Fuselier, \& Burch}]{TurnerDL_2018}
Turner, D.~L., Wilson, L.~B., r., Liu, T.~Z., {et~al.} 2018, Nature, 561, 206, \dodoi{10.1038/s41586-018-0472-9}

\bibitem[{Turner {et~al.}(2021)Turner, Cohen, Michael, Sorathia, Merkin, Mauk, Ukhorskiy, Murphy, Gabrielse, Boyd, Fennell, Blake, Claudepierre, Drozdov, Jaynes, Ripoll, \& Reeves}]{TurnerDL_2021_source}
Turner, D.~L., Cohen, I.~J., Michael, A., {et~al.} 2021, Geophysical Research Letters, 48, \dodoi{10.1029/2021gl095495}

\bibitem[{Uchiyama {et~al.}(2007)Uchiyama, Aharonian, Tanaka, Takahashi, \& Maeda}]{UchiyamaY_2007}
Uchiyama, Y., Aharonian, F.~A., Tanaka, T., Takahashi, T., \& Maeda, Y. 2007, Nature, 449, 576, \dodoi{10.1038/nature06210}

\bibitem[{Vampola \& Korth(1992)}]{VampolaAL_1992}
Vampola, A.~L., \& Korth, A. 1992, Geophysical Research Letters, 19, 625, \dodoi{10.1029/92gl00121}

\bibitem[{Vashenyuk {et~al.}(2011)Vashenyuk, Balabin, \& Gvozdevsky}]{VashenyukEV_2011}
Vashenyuk, E.~V., Balabin, Y.~V., \& Gvozdevsky, B.~B. 2011, Astrophysics and Space Sciences Transactions, 7, 459, \dodoi{10.5194/astra-7-459-2011}

\bibitem[{Vasyliunas(1983)}]{VasyliunasVM_1983}
Vasyliunas, V.~M. 1983, Plasma distribution and flow (Cambridge University Press), 395--453, \dodoi{10.1017/CBO9780511564574.013}

\bibitem[{Vilmer {et~al.}(2003)Vilmer, MacKinnon, Trottet, \& Barat}]{VilmerN_2003}
Vilmer, N., MacKinnon, A.~L., Trottet, G., \& Barat, C. 2003, Astronomy \& Astrophysics, 412, 865, \dodoi{10.1051/0004-6361:20031488}

\bibitem[{Vogt {et~al.}(2010)Vogt, Kivelson, Khurana, Joy, \& Walker}]{VogtMF_2010}
Vogt, M.~F., Kivelson, M.~G., Khurana, K.~K., Joy, S.~P., \& Walker, R.~J. 2010, Journal of Geophysical Research: Space Physics, 115, \dodoi{10.1029/2009ja015098}

\bibitem[{Vogt {et~al.}(2020)Vogt, Connerney, DiBraccio, Wilson, Thomsen, Ebert, Clark, Paranicas, Kurth, Allegrini, Valek, \& Bolton}]{VogtMF_2020}
Vogt, M.~F., Connerney, J. E.~P., DiBraccio, G.~A., {et~al.} 2020, J Geophys Res Space Phys, 125, \dodoi{10.1029/2019ja027486}

\bibitem[{Weisskopf {et~al.}(2000)Weisskopf, Hester, Tennant, Elsner, Schulz, Marshall, Karovska, Nichols, Swartz, Kolodziejczak, \& O'Dell}]{WeisskopfMC_2000}
Weisskopf, M.~C., Hester, J.~J., Tennant, A.~F., {et~al.} 2000, Astrophys J, 536, L81, \dodoi{10.1086/312733}

\bibitem[{Wilson {et~al.}(2016)Wilson, Sibeck, Turner, Osmane, Caprioli, \& Angelopoulos}]{WilsonLB_2016}
Wilson, L.~B., Sibeck, D.~G., Turner, D.~L., {et~al.} 2016, Physical Review Letters, 117, 215101, \dodoi{ARTN 215101 10.1103/PhysRevLett.117.215101}

\bibitem[{Winkler {et~al.}(2003)Winkler, Gupta, \& Long}]{WinklerP_2003}
Winkler, P.~F., Gupta, G., \& Long, K.~S. 2003, The Astrophysical Journal, 585, 324, \dodoi{10.1086/345985}

\bibitem[{Winslow {et~al.}(2013)Winslow, Anderson, Johnson, Slavin, Korth, Purucker, Baker, \& Solomon}]{WinslowRM_2013}
Winslow, R.~M., Anderson, B.~J., Johnson, C.~L., {et~al.} 2013, J Geophys Res-Space, 118, 2213, \dodoi{10.1002/jgra.50237}

\bibitem[{Yaji {et~al.}(2010)Yaji, Tashiro, Isobe, Kino, Asada, Nagai, Koyama, \& Kusunose}]{YajiY_2010}
Yaji, Y., Tashiro, M.~S., Isobe, N., {et~al.} 2010, The Astrophysical Journal, 714, 37, \dodoi{10.1088/0004-637x/714/1/37}

\bibitem[{Yamazaki {et~al.}(2004)Yamazaki, Yoshida, Terasawa, Bamba, \& Koyama}]{YamazakiR_2004}
Yamazaki, R., Yoshida, T., Terasawa, T., Bamba, A., \& Koyama, K. 2004, Astronomy \& Astrophysics, 416, 595, \dodoi{10.1051/0004-6361:20034212}

\bibitem[{Yao {et~al.}(2021)Yao, Fazzini, Chen, Burdonov, Antici, Béard, Bolaños, Ciardi, Diab, Filippov, Kisyov, Lelasseux, Miceli, Moreno, Nastasa, Orlando, Pikuz, Popescu, Revet, Ribeyre, d’Humières, \& Fuchs}]{YaoW_2021}
Yao, W., Fazzini, A., Chen, S.~N., {et~al.} 2021, Nat Phys, 17, 1177, \dodoi{10.1038/s41567-021-01325-w}

\bibitem[{Yeung(2020)}]{YeungPKH_2024}
Yeung, P. K.~H. 2020, Astronomy \& Astrophysics, 640, \dodoi{10.1051/0004-6361/202038166}

\bibitem[{Yuan {et~al.}(2011)Yuan, Liu, Fan, Bi, \& Fryer}]{YuanQiang_2011}
Yuan, Q., Liu, S., Fan, Z., Bi, X., \& Fryer, C.~L. 2011, The Astrophysical Journal, 735, \dodoi{10.1088/0004-637x/735/2/120}

\bibitem[{Zank {et~al.}(2014)Zank, le~Roux, Webb, Dosch, \& Khabarova}]{ZankGP_2014}
Zank, G.~P., le~Roux, J.~A., Webb, G.~M., Dosch, A., \& Khabarova, O. 2014, The Astrophysical Journal, 797, 28, \dodoi{10.1088/0004-637x/797/1/28}

\bibitem[{Zank {et~al.}(2015)Zank, Hunana, Mostafavi, Roux, Li, Webb, Khabarova, Cummings, Stone, \& Decker}]{ZankGP_2015}
Zank, G.~P., Hunana, P., Mostafavi, P., {et~al.} 2015, The Astrophysical Journal, 814, 137, \dodoi{10.1088/0004-637x/814/2/137}

\bibitem[{Zhang {et~al.}(2016)Zhang, Baumjohann, Wang, Dai, \& Tang}]{ZhangLQ_2016}
Zhang, L.~Q., Baumjohann, W., Wang, C., Dai, L., \& Tang, B.~B. 2016, Journal of Geophysical Research: Space Physics, 121, 8773, \dodoi{10.1002/2016ja022397}

\bibitem[{Zhang {et~al.}(2020)Zhang, Chen, Huang, \& Chen}]{ZhangXiao_2020}
Zhang, X., Chen, Y., Huang, J., \& Chen, D. 2020, Monthly Notices of the Royal Astronomical Society, 497, 3477, \dodoi{10.1093/mnras/staa2151}

\end{thebibliography}
\end{document}